\documentclass[aps,prd,superscriptaddress,groupedaddress,nofootinbib,nobibnotes]{revtex4}

\usepackage{graphicx}
\usepackage{dcolumn}
\usepackage{bm}
\usepackage{amssymb}
\usepackage{epstopdf}
\usepackage{amsmath}
\usepackage{amsfonts}
\usepackage{color}
\usepackage{mathrsfs}

\newcommand{\be}{\begin{eqnarray}}

\newcommand{\ee}{\end{eqnarray}}

\newcommand{\ba}{\begin{eqnarray}}
\newcommand{\ea}{\end{eqnarray}}
\newcommand{\nn}{\nonumber}

\newcommand{\bigoh}{\mathcal{O}}

\newcommand{\bea}{\begin{eqnarray}}
\newcommand{\eea}{\end{eqnarray}}
\newcommand{\barr}{\begin{array}}
\newcommand{\earr}{\end{array}}
\def\bal#1\eal{\begin{align}#1\end{align}}

\def\mpl{M_{\rm P}}
\def\k{{\bf k}}

\def\hk{\widehat{\bf k}}

\def\r{{\bf r}}
\def\hr{\widehat{\bf r}}
\def\n{\widehat{\bf n}}

\def\Var{\mbox{Var}}
\def\Cov{\mbox{Cov}}
\def\x{{\bf x}}

\def\hE{{\mathcal{\hat E}}}

\newcommand{\expect}[1]{\left\langle #1 \right\rangle}
\newcommand{\mbf}[1]{\mathbf #1}

\def\ep{\epsilon}

\def\d{\partial}

\def\x{\mbf x}

\def\r{\mbf r}
\def \k {\mbf k}

\begin{document}

\title{Higher N-point function data analysis techniques for heavy particle production and WMAP results}

\author{Moritz M\"unchmeyer}
\affiliation{Perimeter Institute for Theoretical Physics, Waterloo, ON N2L 2Y5, Canada}
\author{Kendrick~M.~Smith}
\affiliation{Perimeter Institute for Theoretical Physics, Waterloo, ON N2L 2Y5, Canada}

\begin{abstract}
We explore data analysis techniques for signatures from heavy particle production during inflation. Heavy particules can be produced by time dependent masses and couplings, which are ubiquitous in string theory. These localized excitations induce curvature perturbations with non-zero correlation functions at all orders. In particular, Ref. \cite{Flauger:2016idt} has shown that the signal-to-noise as a function of the order $N$ of the correlation function can peak for $N$ of order $\mathcal{O}(1)$ to $\mathcal{O}(100)$ for an interesting space of models. As previous non-Gaussianity analyses have focused on $N=\{3,4\}$, in principle this provides an unexplored data analysis window with new discovery potential. We derive estimators for arbitrary N-point functions in this model and discuss their properties and covariances. To lowest order, the heavy particle production phenomenology reduces to a classical Poisson process, which can be implemented as a search for spherically symmetric profiles in the curvature perturbations. We explicitly show how to recover this result from the N-point functions and their estimators. Our focus in this paper is on method development, but we provide an initial data analysis using WMAP data, which illustrates the particularities of higher N-point function searches.

\end{abstract}

\maketitle

%\tableofcontents

%\clearpage
\section{Introduction}

A central goal of modern cosmology is to detect non-Gaussianities in the primordial potential. A detection of such non-Gaussianities would provide us with experimental data on the ultra-high energy physics of the very early universe, at an energy scale far out of reach of even the most futuristic particle colliders. While the primordial potential cannot be directly observed, it seeds the initial perturbations for the matter and radiation in the universe. Currently the most powerful observational technique to reconstruct the primordial potential remains the Cosmic Microwave Background (CMB). In particular the Planck satellite has provided us with the tightest constraints to date on non-Gaussianities of a large number of types \cite{Ade:2015ava,Ade:2015hxq}, without however making a detection. Single-field inflation models have been exhaustively constrained by Planck, to the extent possible with current data. However, once additional degrees of freedom become relevant, a common situation in UV completions of inflation, new forms of non-Gaussianities arise.

In this paper, we develop data analysis methods for a new class of non-Gaussian perturbations arising from heavy fields with time dependent masses \cite{Flauger:2016idt}. These particles have very different observational signatures than those arising from inflaton self-interactions or from additional light degrees of freedom. In general, the probability distribution for the primordial scalar perturbations at some late observation time $t_0$ is given by the functional integral
\be\label{Probdelphi}
P(\zeta)=\int D\chi |\Psi(\zeta, \chi, t_0)|^2 = {\rm tr} (\rho |\zeta\rangle\langle \zeta|)
\ee
where $\rho = \int D\chi |\Psi\rangle\langle\Psi|$ is the density matrix obtained by tracing out any additional fields $\chi$, starting from the state $\Psi (\zeta,\chi, t_0)$ of the full system. In principle, this is a functional's worth of information. In practice, we make progress by reducing this to various simpler observables, such as its moments of fluctuations.  
Most analyses, both in theory and data, have focused on the three-point function (bispectrum) $\langle \Psi | \zeta_{\k_1}\zeta_{\k_2}\zeta_{\k_3} |\Psi\rangle $, or the four-point function (trispectrum) in situations with particular symmetries forbidding the three-point function~\cite{Smith:2015uia}.  
One reason for that is the intuition that perturbative quantum field theory leads to dominant contributions from low-point correlation functions, and with enough symmetry and minimality assumptions the analysis reduces to a few shapes~\cite{Cheung:2007st,Behbahani:2014upa}.          

This approach of constraining a number of bispectra and trispectra however potentially misses models whose signal-to-noise is higher at larger $N$. Indeed, in \cite{Flauger:2016idt} the authors found a regime in which massive particle production generates non-Gaussian $N$-spectra whose signal-to-noise ratio can grow with $N$ for a range of $N$. The existence of these models shows the possibility of controlled inflationary models where the leading signal is not at the lowest non-trivial $N$-point functions and puts in question if we have been exploring our cosmological data with enough generality. In this paper we will analyze specifically the signals predicted in a parametric regime by the models that were developed in~\cite{Flauger:2016idt}. These models are particulary simple in their analytic form, and provide an entrance into high $N$-point function non-Gaussianity search. In practice, in this model, the highest $N$ up to which it makes sense to analyze the data is limited by the probability of observing a particle in our cosmological volume as well as the necessary magnitude of the induced perturbations. We will explain this fact in more detail below.

We start our discussion by deriving an $N$-point function estimator which extends a well-known factorization trick to all $N$. We then derive a resummed estimator which adds the $N$-point estimators in an optimal way. This estimator is our main result. Our estimator requires the $N$-point functions to be in the specific form generated by a Poisson process. In this limit, one can also derive an optimal estimator as a ``profile finder'' in real space, without analyzing non-Gaussianities $N$-by-$N$ in Fourier space. Indeed we show explicitly that our resummed estimator is equivalent to a profile estimator in this limit. However, the full model predictions include interactions between the heavy particles and are naturally evaluated in Fourier space \cite{Flauger:2016idt}. While we do not provide a tractable estimator for such interacting large-$N$ contributions, our factorized analysis demonstrates the behavior of large-$N$ estimators, in particular their non-Gaussian estimator PDFs. In this way we provide an entry into large-$N$ non-Gaussianities, which we hope will spark further investigation of this phenomenology both from a theoretical and an observational point of view.  

Due to the large computation cost of covering the parameter space in \cite{Flauger:2016idt}, as well as the difficulty and novelty of large-$N$ data analysis, we have only performed an initial data analysis using WMAP data \cite{2011ApJS..192...14J}. An analysis using current high-resolution Planck data would however greatly benefit from the experience with WMAP.

The paper is organized as follows. In Sec. \ref{sec:shapes} we review the particle production mechanism and primordial shape function. We present a set of estimators for the simpler case of primordial three dimensional space in Sec. \ref{sec:estimator3d} and discuss their properties. Then we derive equivalent estimators for the realistic case of the CMB perturbations in Sec. \ref{sec:cmb_estimator}. We apply these estimators to WMAP data in Sec. \ref{sec:wmap} and conclude in Sec. \ref{sec:conclusion}. A detailed derivation of our estimator and its properties is given in App. \ref{app:resum} and a review of the classical particle production process can be found in App. \ref{sec:poisson}.

\section{Periodic heavy particle production}
\label{sec:shapes}
In this section we will briefly review the model of periodic heavy particle production, and specify the non-Gaussian shape functions of our analysis. Massive particles are generically produced during inflation if their mass function is time dependent and changes rapidly  enough for non-adiabatic particle production. In this case the particles can be produced even if they are much heavier than the inflationary Hubble scale $H$. In the present case we consider an axion monodromy model \cite{Flauger:2016idt} where the mass term of a heavy field $\chi$ oscillates around a large mean value $\mu$, i.e. the Lagrangian contains the term
\be
\mathcal{L}_m = \frac{1}{2} \chi^2 \left( \mu^2 + 2 g^2 f^2 \cos{\frac{\phi}{f}} \right),
\ee
where $f$ is the axion decay constant, $\phi$ is the inflaton field and $g$ is a coupling constant. This is refered to as ``case b'' in \cite{Flauger:2016idt} (Eq. 2.3). Bursts of particles are produced at each oscillation of the mass function, when the system becomes non-adiabatic. Non-adiabatic particle production is not limited to this specific model. For example it also happens in the case of ``disordered mass functions'' \cite{Amin:2015ftc}, motivated by complicated stochastic UV physics. In addition to causing the creation of particles, the time dependent mass term also results in an oscillating profile of the expanding curvature perturbations associated with each particle. 

The complete primodial $N$-point function for the model under investigation was given in \cite{Flauger:2016idt} in a consistent QFT calculation including interactions. However, a part of these $N$-point functions, which is factorized in momentum space for all $N$, can be understood as a classical Poisson process of particle creation. For a review of this classical picture see App. \ref{sec:poisson}. The additional components generated by the full quantum field theory (including interactions of the produced particles) are hybrid combinations of factorized and resonant shapes (see \cite{Behbahani:2011it,Flauger:2010ja} for a theoretical discussion and \cite{Munchmeyer:2014cca,Meerburg:2015owa} for data analysis techniques). In the present work, we will analyze the factorized component because it is tractable to analyze and dominates the signal-to-noise at least in part of the model parameter space \cite{Flauger:2016idt}.
For the shape under consideration, this factorized contribution to the $N$-point function is \cite{Flauger:2016idt}:
\be\label{Npoints1}
\langle \delta \phi_{\k_1}\dots\delta \phi_{\k_N}\rangle' =  \frac{\bar{n}_\chi}{H^3} H^{N+3} \sum_n(H\eta_n)^{-3}\prod_{i=1}^N \left( \frac{\tilde{c}_b \, h(k_i\eta_n)}{k_i^3} \right) 
\ee
where $\bar{n}_\chi$ is the physical number density, $H$ is the Hubble scale and where 
\be
h(k \eta_n) = \int_{\eta_n}^0 \frac{d\eta'}{\eta'} \sin \left(\frac{\omega}{H} \log \frac{\eta'}{\eta_n} \right) \left[\sin (k \eta') - k \eta' \cos (k \eta') \right].
\ee
The repeated particle production events occur at conformal times
\be
\eta_n = - \frac{1}{H}  e^{\frac{2\pi H}{\omega}(n+\frac{\gamma}{2\pi})}
\ee
with phase $\gamma$. The shape is a function of the frequency $\frac{\omega}{H}$ and the phase $\gamma$. The number of visible events $n$ can be $\mathcal{O}(100)$, depending on the frequency $\omega/H$. In this paper we use curvature perturbations $\zeta = - \frac{H}{\dot{\Phi}} \delta \phi$. Using these variables and $\zeta_{{\rm vac}} = \frac{H^2}{\dot{\Phi}}$ the shape is
\be
\label{Npoints2}
\langle\zeta_{\k_1}\dots\zeta_{\k_N}\rangle' = \bar{n}_\chi \sum_n(H\eta_n)^{-3}\prod_{i=1}^N \left(\frac{- c_b \, h(k_i\eta_n)}{k_i^3} \right) 
\ee
where we defined $c_b = \zeta_{{\rm vac}} \tilde{c}_b$. To get some intution for the amplitude of the profiles compared to vacuum fluctuations, we can use the real space result for a single profile $\zeta(\eta=0,r) \sim \tilde{c}_b \ \zeta_{vac} \  \left(\frac{\omega}{H}\right)^{-1} \sin (\frac{\omega}{H} \log(H r) )$ derived in appendix \ref{sec:realspace}. This equation illustrates the size of the profile for a given $c_b$ and frequency compared to the ``noise'' $\zeta_{vac}$. For some plots of the shape function in the case of the bispectrum see App. \ref{sec:saddlepoint}.

\section{Estimator in primordial space}
\label{sec:estimator3d}

We will first discuss a 3-d estimator in primordial space, which avoids the complications of the CMB but shows the novel features of this non-Gaussian shape. We then adapt this estimator to the CMB in the next section. For simplicity, here we drop the sum over events and define $h(k)$ so that
\be
\label{3destim1}
\langle \zeta_{\k_1} \cdots \zeta_{\k_N} \rangle'_c = \alpha \prod_{i=1}^N h(k_i)
\ee
We would like to find an estimator for $\alpha$. Note that $\alpha$ is a number density with units $[1/L^3]$ (rather than an dimensionless non-Gaussianity amplitude).

\subsection{Arbitrary N estimator and resummation}

Consider the following $N$-point estimator:
\ba
\hE_N = \frac{1}{N!} \int_{\k_1\cdots\k_N} \frac{\langle\zeta_{\k_1} \cdots \zeta_{\k_N} \rangle^*}{P(k_1) \cdots P(k_N)} \zeta_{\k_1} \cdots \zeta_{\k_N}. \label{eq:en_def_3d}
\ea
Following a standard recipe, each $N$-tuple $(\k_1,\cdots,\k_N)$ is weighted proportionally to
the $N$-point signal $\langle \zeta_{\k_1} \cdots \zeta_{\k_N} \rangle$, multiplied by the inverse 
covariance $P(k_1)^{-1} \cdots P(k_N)^{-1}$.
Note that $\hE_N$ is not quite the optimal estimator for the $N$-point signal $\langle \zeta_{\k_1} \cdots \zeta_{\k_N} \rangle$,
since the optimal estimator would include terms of orders $(N-2), (N-4), \cdots$ (e.g.~\cite{Smith:2015uia}).
We will shortly define a resummed estimator that self-consistently includes all of these terms, and also
combines $N$-point signals $\langle \zeta_{\k_1} \cdots \zeta_{\k_N} \rangle$ from different values of $N$.

Using the usual factorization trick, first used for non-Gaussianities in the bispectrum in \cite{Komatsu:2003iq}, we can write the estimator $\hE_N$ in real space as:
\be
\hE_N = \frac{1}{N!} \int d^3\r \, \psi(\r)^N
\ee
where we define the field $\psi$ by filtering $\zeta$ in Fourier space:
\be
\psi_\k = \frac{h(k)}{P(k)} \zeta_\k  \label{eq:psi_def_3d}
\ee
Interestingly, the computational cost of the $N$-point estimator is almost the same as the one for the three-point function.
We will refer to $\hE_N$ as the \textit{$N$-point estimator}. %\kms{Suggest ``$N$-point estimator'' here and elsewhere?}

What is the optimal estimator $\hE_{\rm opt}$ for the parameter $\alpha$ in Eq.~(\ref{3destim1})?
As an ansatz, we consider an arbitrary linear combination $\hE_{\rm opt} = \sum_N W_N \hE_N$\
of $N$-point estimators, in order to combine information from different values of $N$.
Then we can ask, which choice of coefficients $W_N$ minimizes the variance $\Var(\hE_{\rm opt})$,
subject to the constraint that the estimator is unbiased, i.e.~$\langle \hE_{\rm opt} \rangle = \alpha$
to first order in $\alpha$.

In Appendix~\ref{app:resum}, we solve this constrained minimization problem, and show that the
result can be ``resummed'' to write $\hE_{\rm opt}$ in the following compact form:
\be
\hE_{\rm opt} = F_{\rm tot}^{-1} \int d^3\x \, \Big( e^{-\langle\psi^2\rangle/2} e^{\psi(\x)} - 1 \Big)  \label{eq:eopt_3d}
\ee
Here, $\langle\psi^2\rangle = \int dk \, (k^2/2\pi^2) h(k)^2 / P(k)$ is the real-space variance of the $\psi$ field.
The prefactor $F_{\rm tot}$ is given explicitly in Eq.~(\ref{eq:Ftot_3d}).
We refer to $\hE_{\rm opt}$ as the \textit{resummed estimator}.

\subsection{Relation to profile finding}
\label{sec:profilefinding}
The optimal estimator $\hE_{\rm opt}$ has an interesting reinterpretation as a profile-finding statistic.
Given a realization of the 3-d field $\zeta$, consider two hypotheses.
The null hypothesis $H_0$ is that $\zeta$ is a Gaussian random field.
Hypothesis $H_1$ is that a random profile has been added to a Gaussian field:
\be
\zeta(\x) = \zeta_G(\x) + h(\x-\x_0)
\hspace{1cm}
\mbox{where } h(r) = \int \frac{d^3\k}{(2\pi)^3} h(k) e^{i\k\cdot\r}
\ee
where $\zeta_G$ is Gaussian, and the profile center $\x_0$ is uniform random.

The optimal statistic for distinguishing these hypotheses is the likelihood ratio:
\ba
\hE = \frac{\mathcal{L}(\zeta|H_1)}{\mathcal{L}(\zeta|H_0)}
\ea
and the conditional likelihoods which appear are given by
\ba
\mathcal{L}(\zeta|H_0) &=& \exp\left[ -\frac{1}{2} \int \frac{d^3\k}{(2\pi)^3} \frac{|\zeta(\k)|^2}{P(k)} \right] \\
\mathcal{L}(\zeta|H_1) &=& V^{-1} \int d^3\x_0 \, \exp \left[ -\frac{1}{2} \int \frac{d^3\k}{(2\pi)^3} \frac{|\zeta(\k) - h(k) e^{-i \k \cdot \x_0}|^2}{P(k)} \right]
\ea
where $V$ is the 3-d volume (assumed finite).
Plugging these in, the likelihood ratio statistic is:
\ba
\hE
&=& V^{-1} \int d^3\x_0 \, \exp\left[ \int \frac{d^3\k}{(2\pi)^3} \left( \frac{h(k)}{P(k)} \zeta(\k) e^{i\k\cdot\x_0} - \frac{h(k)^2}{2 P(k)} \right) \right] \nn \\
&=& V^{-1} \int d^3\x_0 \, \exp\left[ \psi(\x_0) - \frac{\langle\psi^2\rangle}{2} \right]
\ea
where $\psi(\x)$ was defined in Eq.~(\ref{eq:psi_def_3d}).
Comparing with Eq.~(\ref{eq:eopt_3d}), we reproduce the result of the resummed estimator in the previous section,
up to overall normalization and an additive constant.
It is clear from this formulation that this likelihood is not correct when several overlapping profiles are present,
which is equivalent to the first order in $\alpha$ condition in our resummed estimator (see Appendix~\ref{app:resum}). 

\section{Estimator for the CMB}
\label{sec:cmb_estimator}

In this section we translate our estimators from primordial space to the CMB. We also remove the notational simplifications of the previous section and consider the full shape function in Eq.~(\ref{Npoints2}), including the sum over events, 
\be\label{Npoints}
\langle\zeta_{\k_1}\dots\zeta_{\k_N}\rangle_c = M \sum_n \eta_n^{-3} \left( \prod_{i=1}^N \frac{- c_b \, h(k_i\eta_n)}{k_i^3} \right) \, (2\pi)^3 \delta^3\Big( \sum\k_i \Big).
\ee
Here we have defined the dimensionless number $M = \frac{\bar{n}}{H^3}$, where $\bar{n}$ is the physical number density of particles produced in each event $n$. We want to estimate this dimensionless amplitude $M$ (rather than the density $\bar{n}$) as a function of profile amplitude $c_b$. The primordial physics interpretation of $M$ is that it is the average number of particles produced in an inflationary Hubble volume at each production event $n$. For data analysis it is more convenient to think of $M$ as the average number of particles per profile volume, which is invariant under the expansion of the universe. This interpretation is valid because the comoving size of the profiles after horizon exit is $R=-\eta_n$, as we review in App. \ref{sec:realspace}. The comoving number density of particles produced at time $\eta_n$ is $\frac{\bar{n}}{(- \eta_n H)^3}$,  i.e. we will see the largest number of profiles from the latest visible production times, and the comoving profile volume is $R^3=(-\eta_n)^3$.

Profiles are visible in the data if their size is at least of order the pixel size of the experiment. For our CMB experiment, the number of pixels is of order $\ell_{max}^2 \sim 10^6$.
It follows that we cannot constrain $M$ to be smaller than $M \sim \ell_{max}^{-2}$, since the number of observable particles per CMB sky
would be $\lesssim 1$.
We will get back to this subtlety when we interpret our results in Sec. \ref{sec:wmap}.

\subsection{CMB $N$-point function}

To calculate the CMB N-point function from the primordial N-point function, one applies the usual projection formula
\be
a_{\ell m} = 4\pi i^\ell \int \frac{d^3\k}{(2\pi)^3} \Delta_\ell(k) \zeta_\k Y_{\ell m}^*(\hat{\bf{k}})
\ee
and obtains:
\be
\big\langle a_{l_1m_1} \cdots a_{l_Nm_N} \big\rangle_c =
\sum_n f_n
  \left( \prod_{i=1}^N  4\pi i^{\ell_i}   \int \frac{d^3\k_i}{(2\pi)^{3}} \, 
  \Delta_{\ell_i}(k_i) S_n(k_i) Y_{\ell_im_i}^*(\hk_i) \right)
  (2\pi)^3 \delta^3\Big(\sum k_i \Big)
\ee
where we have defined
$f_n = M \eta_n^{-3}$ and $S_n(k) = - c_b h(k\eta_n)/k^3$.
The delta function can be represented as usual as an exponential that is then Rayleigh expanded, i.e. 
\be
(2\pi)^3 \delta^3(\sum\k_i) = \int d^3\r\, \prod_{i=1}^N \left( 4\pi
\sum_{\ell'_i m'_i} i^{\ell'_i} j_{\ell'_i}(k_ir) Y_{\ell'_i m'_i}^*(\hr) Y_{\ell'_i m'_i}(\hk_i) \right)
\ea
Plugging in and performing the angular integration over $\k$ one obtains
\ba
\big\langle a_{l_1m_1} \cdots a_{l_Nm_N} \big\rangle_c
&=&
\sum_n f_n \int d^3\r \, \prod_{i=1}^N  \frac{ 2 \ d k_i \ k_i^2}{\pi} \, 
      \bigg( \Delta_{\ell_i}(k_i) j_\ell(k_ir)  Y_{\ell_im_i}^*(\hr) \bigg) S_n(k_i) \\
&=&
  \sum_n f_n \int d^3\r \, \bigg( \prod_{i=1}^N M^{n}_{\ell_i}(r)  Y_{\ell_i m_i}^*(\hr) \bigg)  \label{eq:cmb_npoint_signal}
\ea
where we have defined
\be
M^{n}_{\ell}(r) &=& \frac{2}{\pi} \int dk \ k_i^2 \, \Delta_{\ell}(k) j_\ell(k r) S_n(k)  
\ee

\subsection{CMB $N$-point estimator and resummation}

The \textit{$N$-point CMB estimator} for a general $N$-point signal $\langle a_{l_1m_1} \cdots a_{l_Nm_N} \rangle_c$ is:
\be
\hE_N = \frac{1}{N!} \sum_{\ell_i m_i} \big\langle a_{l_1m_1} \cdots a_{l_Nm_N} \big\rangle_c^*
   \Big[ (C^{-1}a)_{\ell_1m_1} (C^{-1}a)_{\ell_2m_2} \dotsc (C^{-1}a)_{\ell_Nm_N}  \Big]
\ee
where $C^{-1} = (S+N)^{-1}$ is the inverse signal + noise covariance.
This expression applies for an arbitrary noise model, parameterized by the noise covariance matrix $N$.
Plugging in the $N$-point signal in Eq.~(\ref{eq:cmb_npoint_signal}), one obtains
\be
\label{eq:cmbestimator1}
\hE_N = \frac{1}{N!} \sum_n f_n \int d^3\r \, \psi_n(\r)^N
\ee
where
\be
\psi_n(\r) = \sum_{lm} M^n_l(r) \, \big(C^{-1}a\big)_{lm} \, Y_{lm}(\hr)
\ee
As was the case for the primordial space estimator in Eq.~(\ref{eq:en_def_3d}), this estimator is not optimal, as it does not take into account corrections from lower $N$-point functions and masking. We expect corrections from masking to be small, since the present shape is dominated by equilateral $N$-point configurations, rather than squeezed configurations, which are much more sensitive to the presence of a mask. To build intuition, we will apply this naive estimator to the data, in addition to the optimal estimator to be defined shortly. Because the amplitude of the $N$-point function estimator is not physically normalized, due to the presence of the correction terms from lower $N$-point function, we will only present results divided by the square root of the estimator variance, i.e. we define
\be
\label{eq:gausssigmas}
\tilde{\mathcal{E}}_{\omega_i,\phi_i}^N = \frac{\mathcal{E}_{\omega_i,\phi_i}^N}{ \sqrt{\mbox{Var}(\mathcal{E}^N_{\omega_i,\phi_i})} }. 
\ee
It should be noted that these estimates are not ``sigmas'' because the estimator PDF is significantly non-Gaussian for $N>4$. Further, we will only use the $N$-point function estimator for odd N for which $\langle \hE^N \rangle_{G} = 0$.

To deal with these shortcomings, as in the case of the primordial space estimator, in Appendix~\ref{app:resum} we derive the optimal \textit{resummed CMB estimator} $\hE_{\rm opt}$, combining $N$-point functions from
different values of $N$.  The final result can be written in the resummed form:
\be
\hE_{\rm opt} = F^{-1} \sum_n f_n \int d^3\r \, \Big( e^{-\langle\psi_n(\r)^2\rangle/2} e^{\psi_n(\r)} - 1 \Big) \label{eq:eopt_2d}
\ee
where the overall prefactor $F$ is given in Eq.~(\ref{eq:F_2d}).
In the special case of a diagonal covariance matrix, the quantity $\langle \psi_n(\r)^2 \rangle$ is given by:
\be
\langle \psi_n(\r)^2 \rangle = \sum_l \frac{2l+1}{4\pi} \frac{M_l^n(r)^2}{C_l} 
\ee
We find that the resummed estimator involves a simple change of normalisation of the estimator amplitudes, in the approximation of a diagonal covariance matrix. Note that the resummed estimator~(\ref{eq:eopt_2d}) self-consistently includes all correction terms for an arbitrary noise model. We have however made the approximation of a diagonal covariance matrix, which should be good for our Wiener filtered maps, to avoid the computational cost of evaluating $\langle \psi_n(\r)^2 \rangle$ from Monte Carlo.

\section{Results from WMAP}
\label{sec:wmap}

In this section we apply our estimators to real data from the WMAP satellite. We use WMAP7 \cite{2011ApJS..192...14J} data with $\ell_{\rm max}=1000$, as well as a set of 40 equivalent Gaussian simulations for Monte Carlo estimates. Both the data map and the simulations have been Wiener filtered using the conjugate gradient with multigrid preconditioner from \cite{Smith:2007rg}.

The shape under consideration has two free parameters, the frequency $\omega/H$ and phase $\gamma$ (contained in $\eta_n$). We have to scan over both of these parameters sufficiently tightly to not miss a potential detection. We sample the phase with 12 values of $\gamma \in (0,2\pi)$. We analyze the frequency range $1 < \omega/H < 100$ (which covers a large part of the theoretical parameter space given as $\frac{1}{10} < \omega/H < 200$), and found that we require about 100 sampling points to cover the full frequency space. We need to know the range of the event sum $\eta_n$ which is observable in the CMB. The $k$ range in which the CMB transfer functions contribute to CMB scales $2 \leq \ell \leq 1000$ (WMAP resolution) is $k_{{\rm min}} \simeq 10^{-5} \, \mathrm{Mpc}^{-1}$ to $k_{{\rm max}} \simeq 0.1 \, \mathrm{Mpc}^{-1}$. At the largest frequency $\omega / H = 100$ at WMAP resolution one needs to calculate about 200 events. In addition to the sum over events, one has to sample the $r$ integral of the CMB estimator with about 100 sampling points. Covering the parameter space with the CMB estimator thus requires a large number of spherical harmonics transforms, and is computationally very expensive. For this reason for our initial analysis we have restricted ourselves to WMAP resolution (as opposed to using more high resolution Planck CMB data) and analyzed only enough Monte Carlo maps to prove non-significance (here 40 maps). Our main focus is on demonstrating the properties of the estimators on data.

\subsection{Results for $N$-point estimators}

\begin{figure}
\resizebox{0.9\hsize}{!}{
\includegraphics{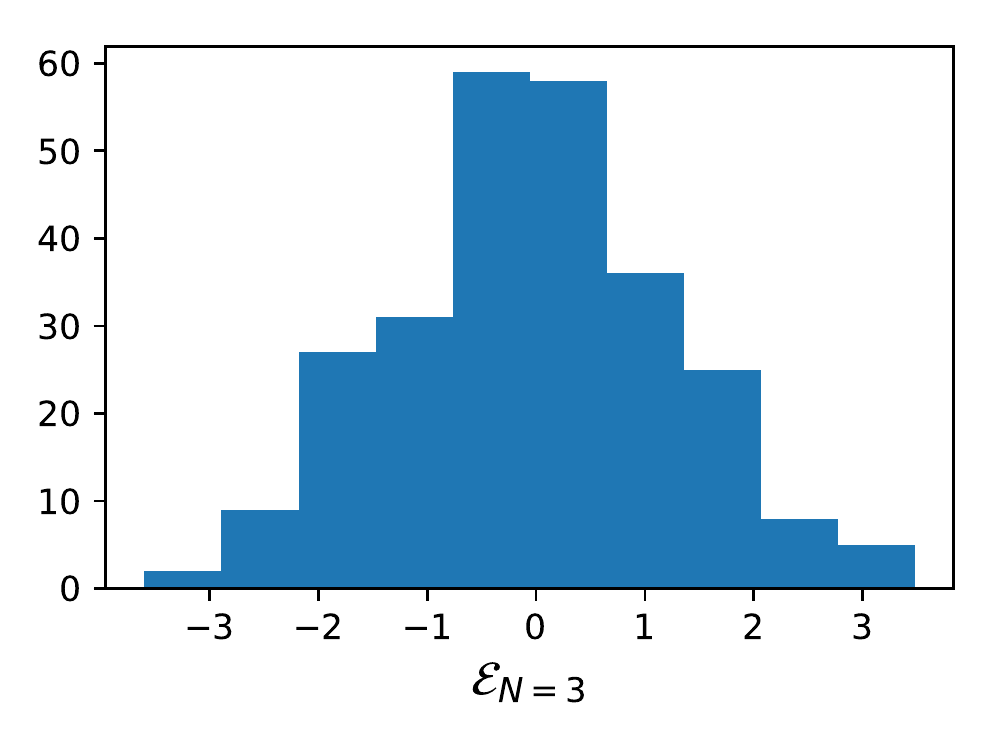}\includegraphics{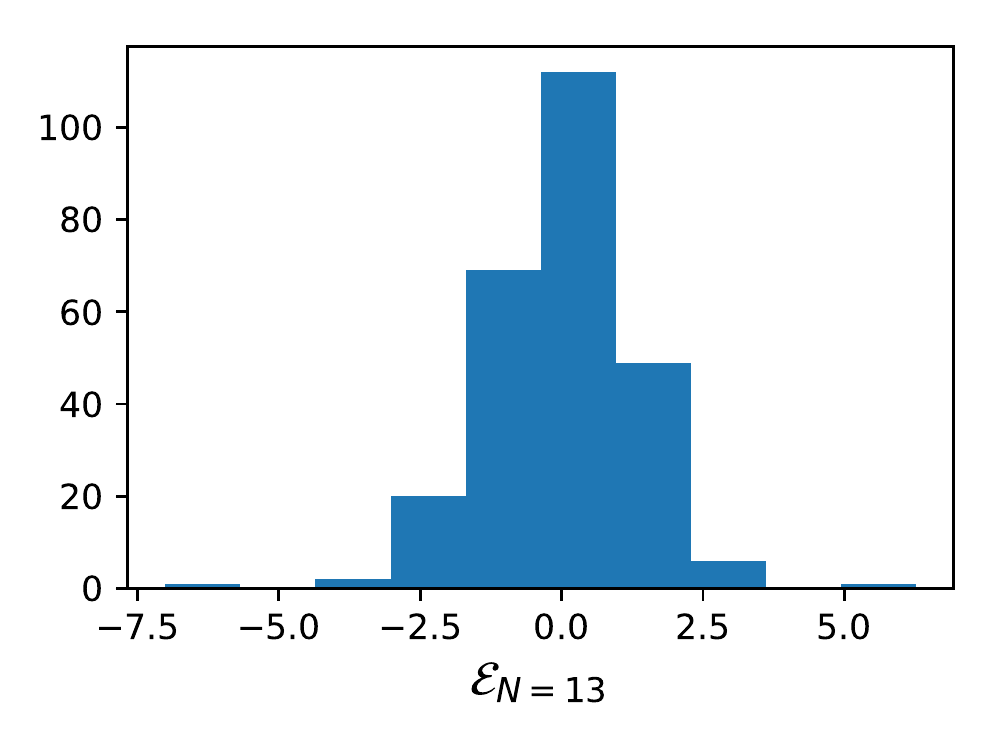}\includegraphics{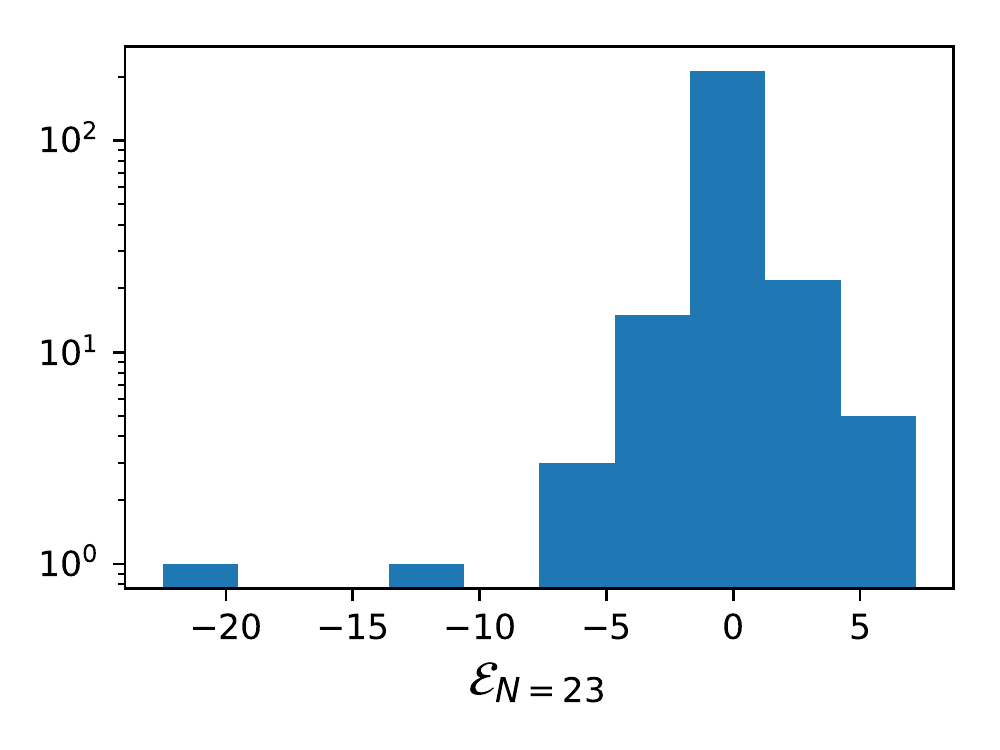}
}
\caption{Histogram of the normalized N-point CMB estimator Eq. (\ref{eq:gausssigmas}) for Gaussian CMB maps for three different values of $N$. For $N=3$ (left) we see the familiar Gaussian form. For larger $N$ the tails become successively larger. Note that the plot for $N=23$ (right) is in log scale. The plots were generated from Monte Carlo maps using several estimator frequencies around $\omega/H=40$. }
\label{fig:histos}
\end{figure}

In this section we apply the $N$-point estimator in Eq. \eqref{eq:cmbestimator1} to the WMAP data and Gaussian simulations.
%\subsubsection{Non-Gaussianity of the estimator PDF and peak statistic}
The PDF of our $N$-point function estimators is non-Gaussian, even for an underlying Gaussian map. To see why this is the case intuitively recall that the $N$-point estimator is of form
\be
\mathcal{E}_N \propto \int d^3\r (\psi(\r))^N
\ee
where $\psi(\r)$ is the filtered map. For large $N$ the integral will be dominated by small outlier regions in the map. While for $N=3$ the bispectrum estimator has the familiar Gaussian form, estimators for $N \geq 5$ visibly deviate from a Gaussian. We illustrate this behavior in Fig.~\ref{fig:histos} with Monte Carlo CMB maps. Unfortunately even for an ideal full sky experiment with uniform coverage and no noise it appears to be difficult to find an analytic expression for the estimator PDF, which could be used to calculate frequentist significances. We thus use Gaussian simulations to estimate the significance without analytic knowledge of the estimator PDF.

To obtain a frequentist significance we look for the most significant peak in the space of frequencies and phases at each $N$, i.e. we define the \textit{peak statistic}
\be
\label{eq:peakestimator}
\mathcal{E}_{\rm peak}^N = \max_{\omega_i,\phi_j} |\tilde{\mathcal{E}}_{\omega_i,\phi_i}^N|.
\ee
We then compare the value of the peak statistic that we find in the data to that from Gaussian simulations to, in principle, obtain a p-value at each $N$. In practice, to save computation, we do not run enough simulations to calculate a precise p-value but only rank the amplitude in the data with respect to those in Gaussian simulations. For the peak statistic to cover the frequency and phase parameter space with equal weight, the estimators need to have the same PDF for each sampling point. For this reason we have used the inverse variance weighted estimators $\tilde{\mathcal{E}}_{\omega_i,\phi_i}^N$ defined in Eq.~(\ref{eq:gausssigmas}).

\begin{figure}
\resizebox{0.5\hsize}{!}{
\includegraphics{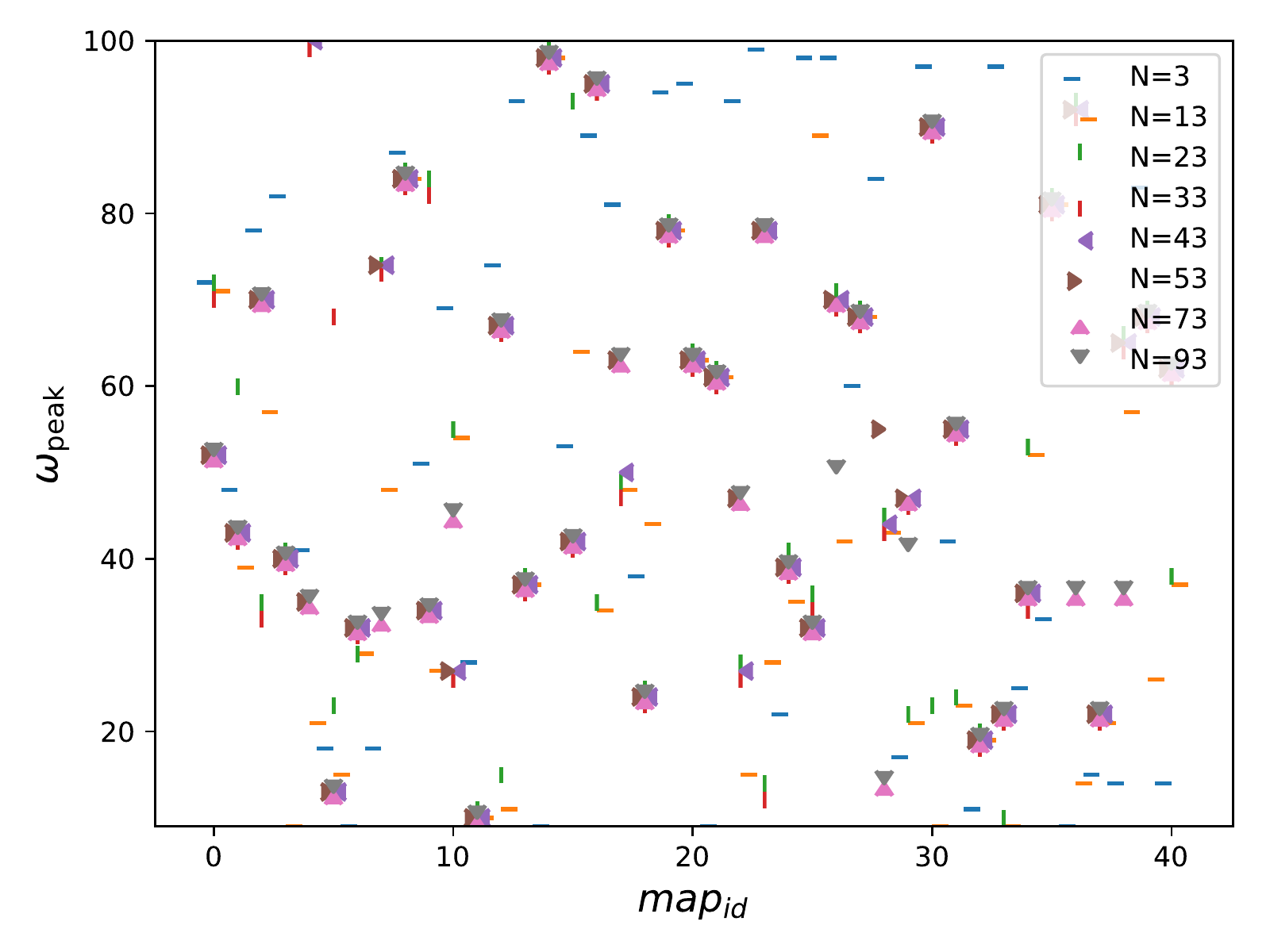}
}
\caption{Scattering of the peak frequencies of $\tilde{\mathcal{E}}_{\omega_i,\phi_i}^N$  for 40 Gaussian Monte Carlo maps and different $N$. The peak frequencies appear randomly distributed as they should be for Gaussian maps. The plot also illustrates that large $N$ are very correlated.}
\label{fig:res3}
\end{figure}

We have obtained the estimator variance for this weighting from our simulations. We have found that the large tails of the estimator distribution make the variance estimate converge slowly. However it turns out that the estimator variance is quite smooth in $\omega$ and $\phi$, i.e. it varies slowly with these parameters. Therefore one can use the estimates of $F$ at different nearby frequency points to reduce the number of neccessary maps to make the variance estimate converge. In praxis we have smoothed the estimated function $F^N(\omega,\phi)^{1/N}$ with a Savitzky-Golay filter. We checked empirically that this procedure leads to an even distribution of peak frequencies in simulations. Fig.~\ref{fig:res3} shows a scatter plot of the peak frequency of $\tilde{\mathcal{E}}_{\omega_i,\phi_i}^N$ drawn from Gaussian simulations. We see that the peak frequencies appear randomly distributed over $\omega$ as they should be.

\begin{figure}
\resizebox{0.8\hsize}{!}{
\includegraphics{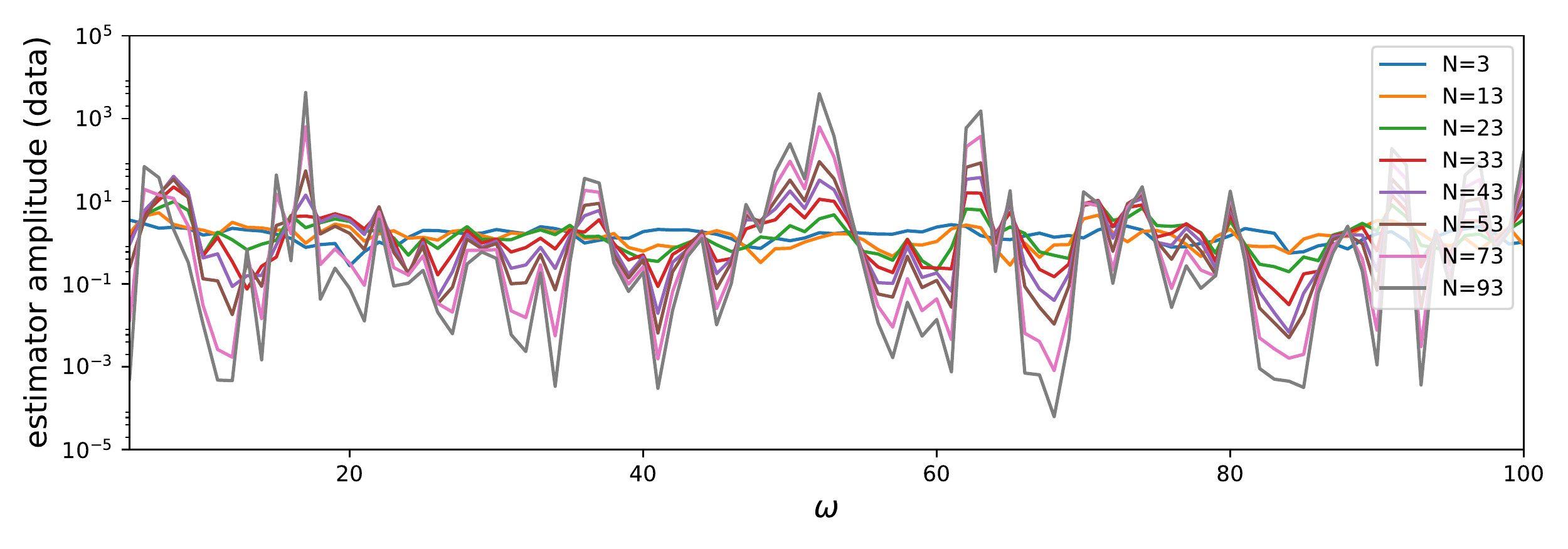}
}
\resizebox{0.8\hsize}{!}{
\includegraphics{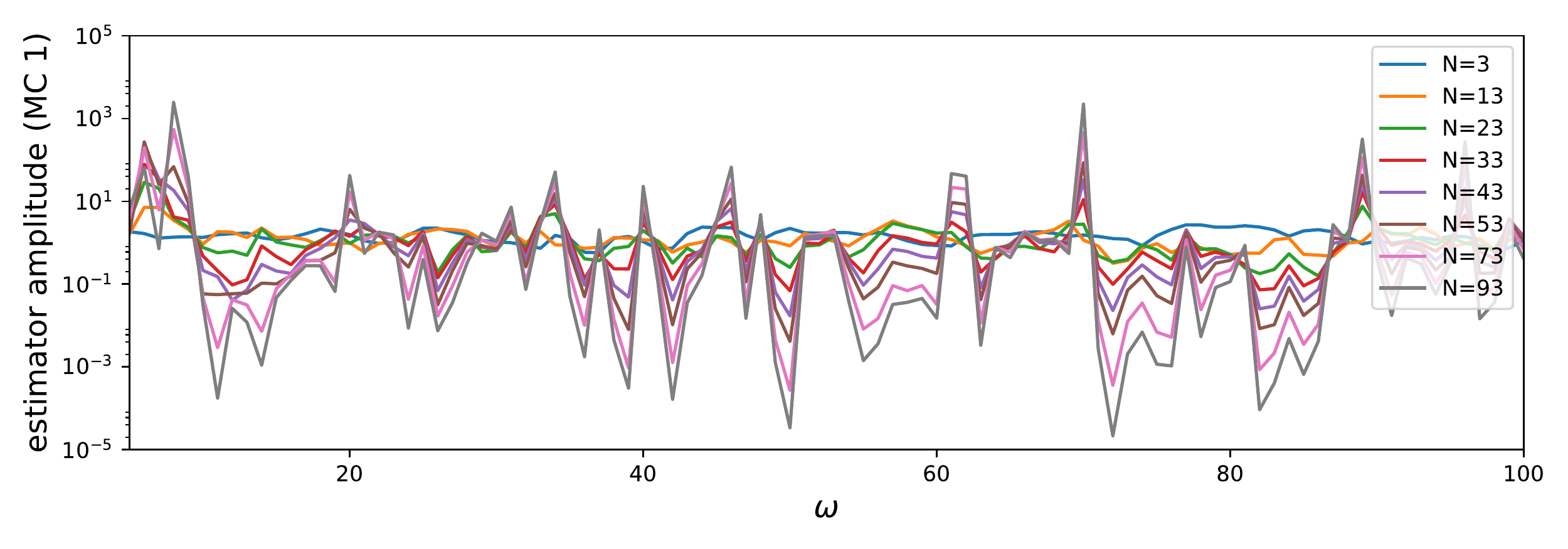}
}
\resizebox{0.8\hsize}{!}{
\includegraphics{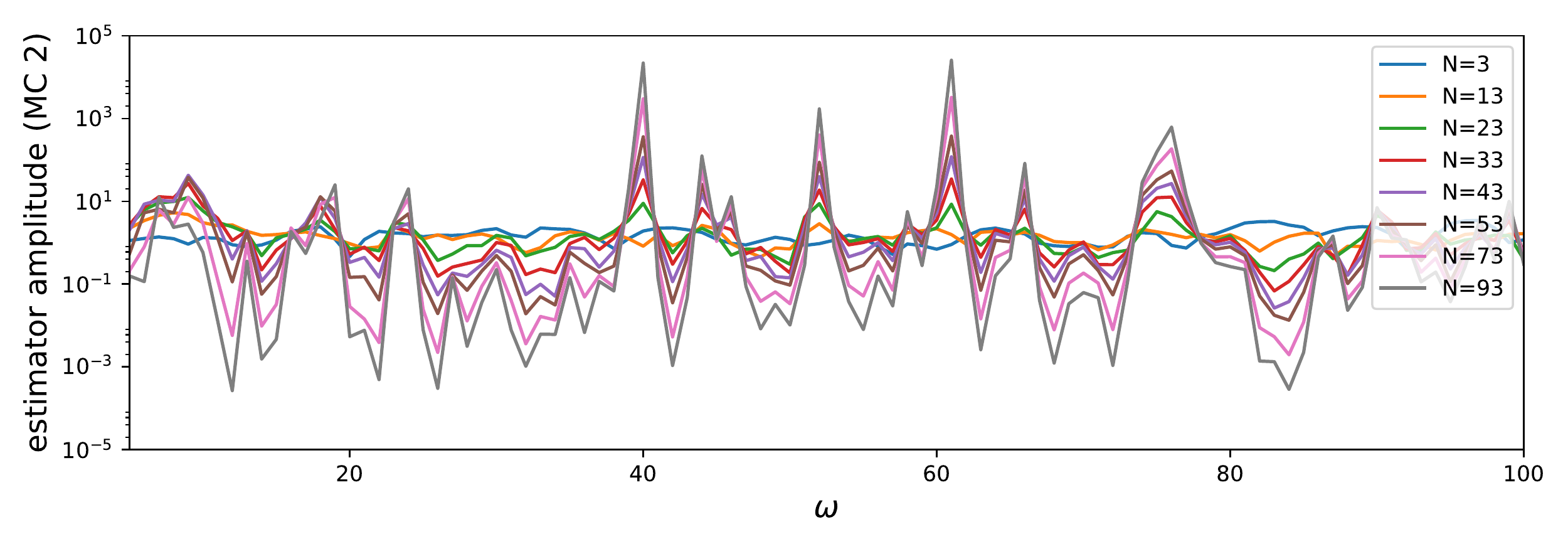}
}
\caption{Estimator amplitudes $\tilde{\mathcal{E}}_{\omega}^N$ (Eq.~\eqref{eq:gausssigmas}) for WMAP (top) and two Gaussian Monte Carlo maps, as a function of frequency for eight different values of $N$. The estimator for each $\omega$ has been maximized over $\phi$. There are no unusually high peaks in the WMAP data in this frequency range. We also see that the estimates at large $N$ are very correlated.}
\label{fig:res1}
\end{figure}

\begin{figure}
\resizebox{0.38\hsize}{!}{
\includegraphics{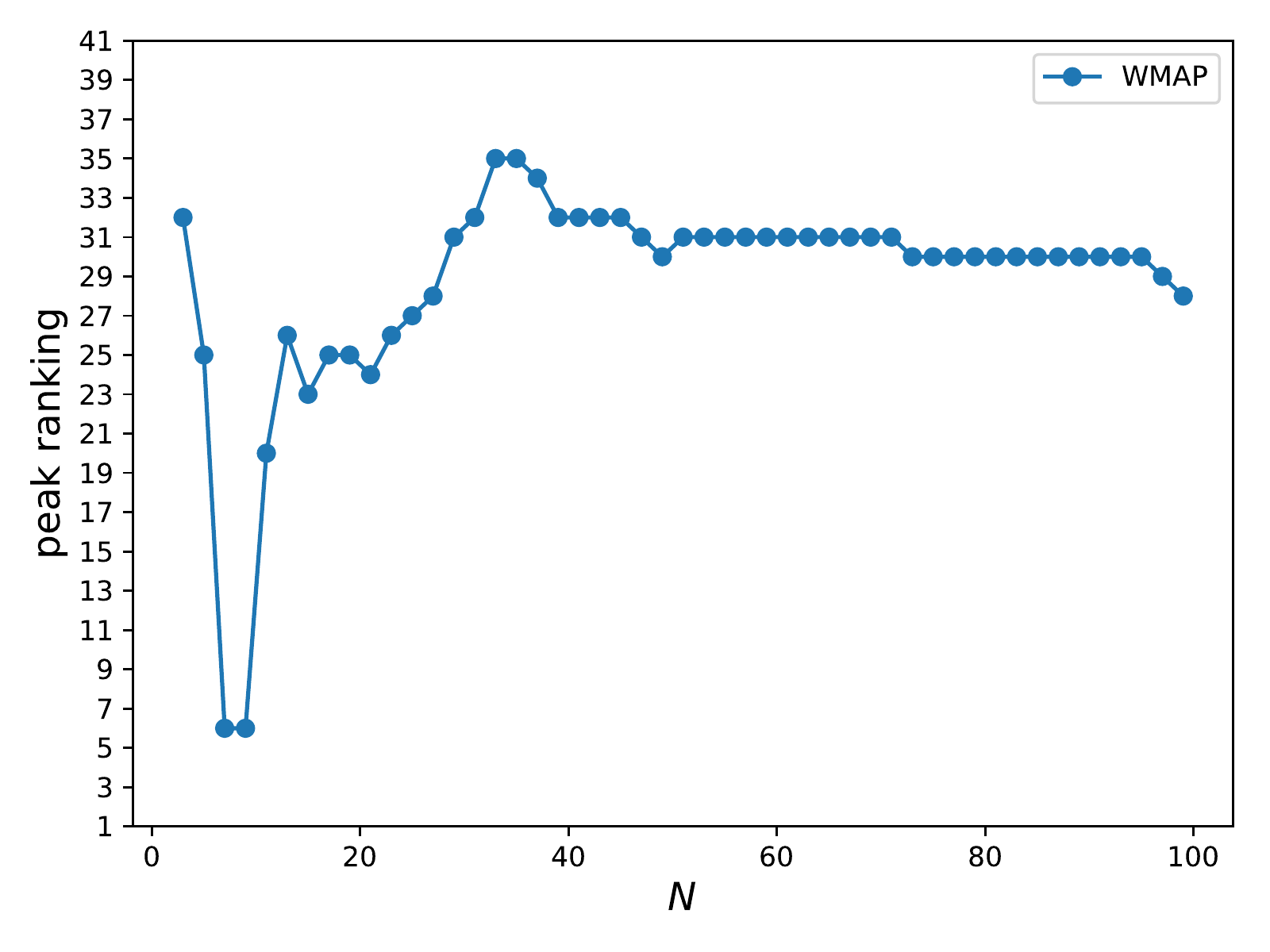}
}
\caption{Left: Peak ranking of the WMAP 7 data compared with 40 Monte Carlo maps as a function of $N$ (every odd $N\geq 3$). Here rank 1 means the highest peak in the ensemble of maps and rank 41 the lowest. The WMAP map does not stand out of the Monte Carlo maps.}
\label{fig:res2}
\end{figure}

We show results for the estimator $\tilde{\mathcal{E}}_{\omega_i,\phi_i}^N$ in Fig.~\ref{fig:res1}, for the data and three MC maps. This quantity, defined in Eq.~\eqref{eq:gausssigmas} is normalized so that it is independent of the overall normalization of the shape function, and for low $N$ (with Gaussian PDF) can be interpreted as the number of sigmas. It gives an impression of the local size of the peaks, and shows that there is no immediately significant peak in the data, compared to the Monte Carlo examples. In Fig.~\ref{fig:res2} we rank the result with respect to MC maps, with the peak estimator defined in Eq.~\eqref{eq:peakestimator}. Again we find no extraordinary value in the data, and are fully consistent with Gaussianity. The plots also illustrate the high correlation between different $N$, in particular at large $N$.

\subsection{Results for the resummed estimator}

\begin{figure}
\resizebox{0.9\hsize}{!}{
\includegraphics{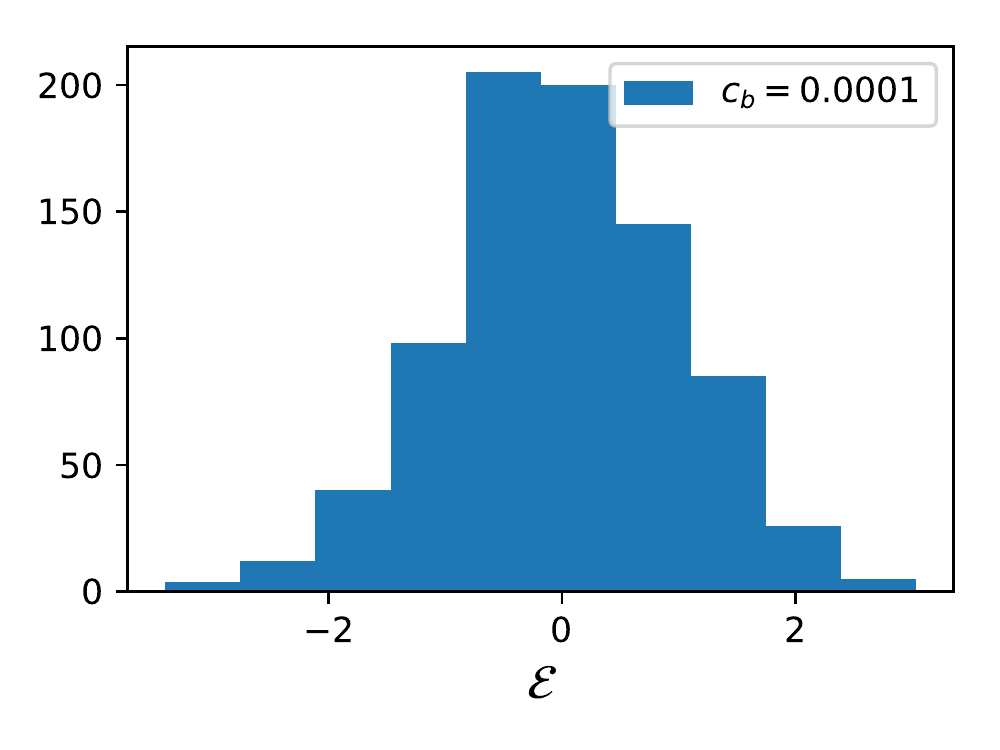}\includegraphics{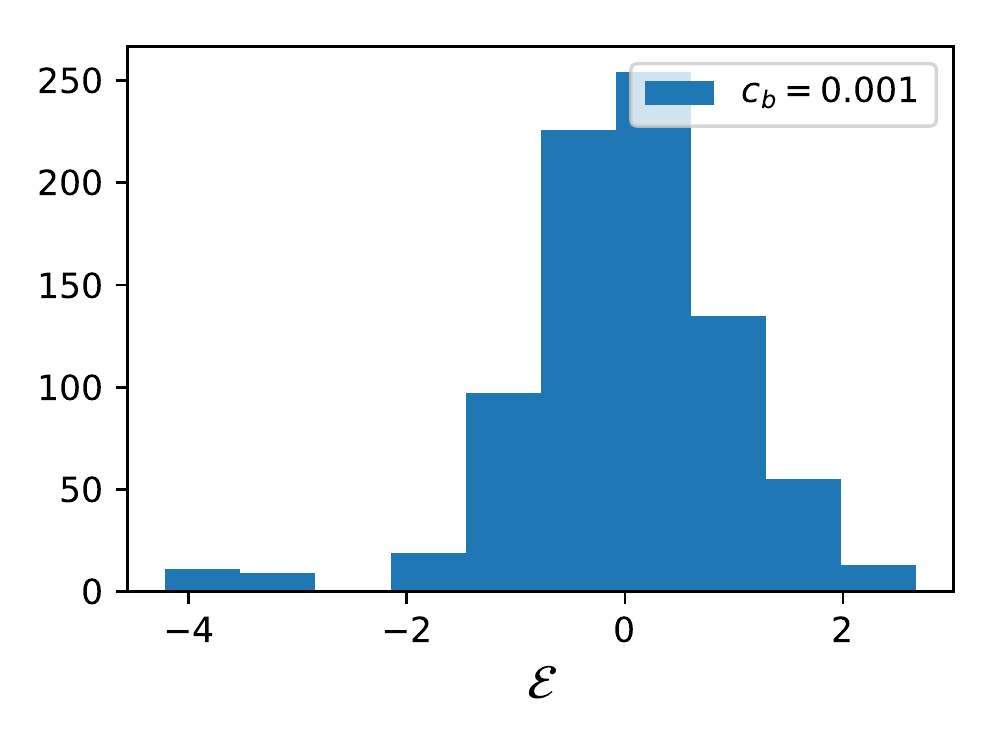}\includegraphics{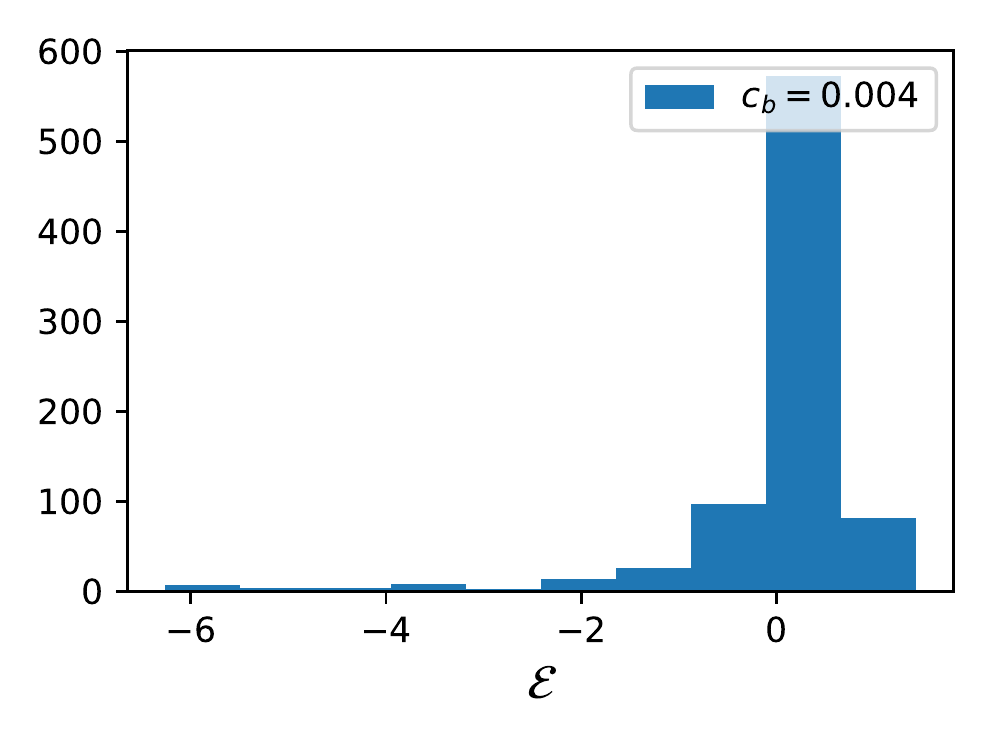}
}
\caption{Histogram of the resummed estimator (normalized by the variance) of Gaussian CMB maps for three different values of $c_b$ (0.0001 left, 0.001 middle, 0.004 right). The plots were generated from Monte Carlo maps with estimator frequencies around $\omega/H=40$. The mean was subtracted. For low $c_b$ we find an approximately Gaussian distribution as expected. } 
\label{fig:histosEXP}
\end{figure}

Now we evaluate the resummed estimator $\mathcal{E}^{\omega,\phi}_{\rm opt}$  defined in Eq.~\eqref{eq:eopt_2d}. The shape function is normalized as in Eq.~\eqref{Npoints}, with the goal to estimate $M$. 
To normalize the estimator, we estimated the Fisher matrix from the variance of the estimator on Gaussian Monte Carlo maps.
We first inspect the PDF of the resummed estimator, which is shown by the histograms in Fig. \ref{fig:histosEXP}, drawn from Gaussian Monte Carlo realisations. As expected we find a non-Gaussian estimator PDF for large $c_b$.

\begin{figure}
\resizebox{0.7\hsize}{!}{
\includegraphics{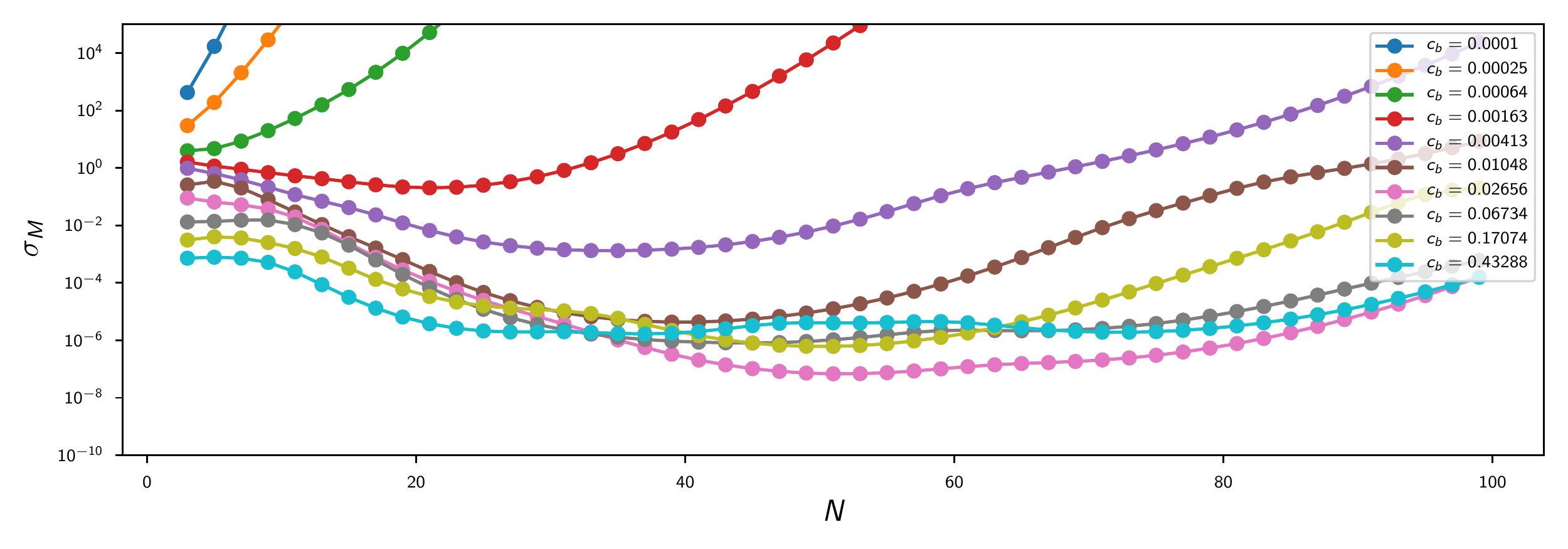}
}
\caption{Sensitivity $\sigma_{M}:=1/\sqrt{F}$ values, for different values of $c_b$, at the example of $\omega/H=30$, as a function of N. We extracted the N-by-N terms from the resummed exponential estimator and calculated the Fisher matrix from the variance over 40 Monte Carlo maps. The lower $\sigma$, the better the sensitivity. We see that in general for larger $c_b$ the main contribution comes from larger $N$. The plot shows all odd N starting at $N=3$. Note that the variance used to estimate F is not Gaussian, so the interpretation as ``sigmas'' does not hold.}
\label{fig:res4}
\end{figure}

We are also interested in which values of $c_b$ get the largest signal-to-noise contribution from which $N$. This is illustrated in Fig.~\ref{fig:res4}. There is an optimal value in $N$ for a given $c_b$. For $c_b$ values below $0.001$ we find that the bispectrum is the dominant contribution to the non-Gaussianity. Note that the $\sigma_{NL}$ plotted here does not have the usual Gaussian interpretation at higher $N$, due to the non-Gaussianity of the estimator PDF. 

\begin{figure}
\resizebox{0.7\hsize}{!}{
\includegraphics{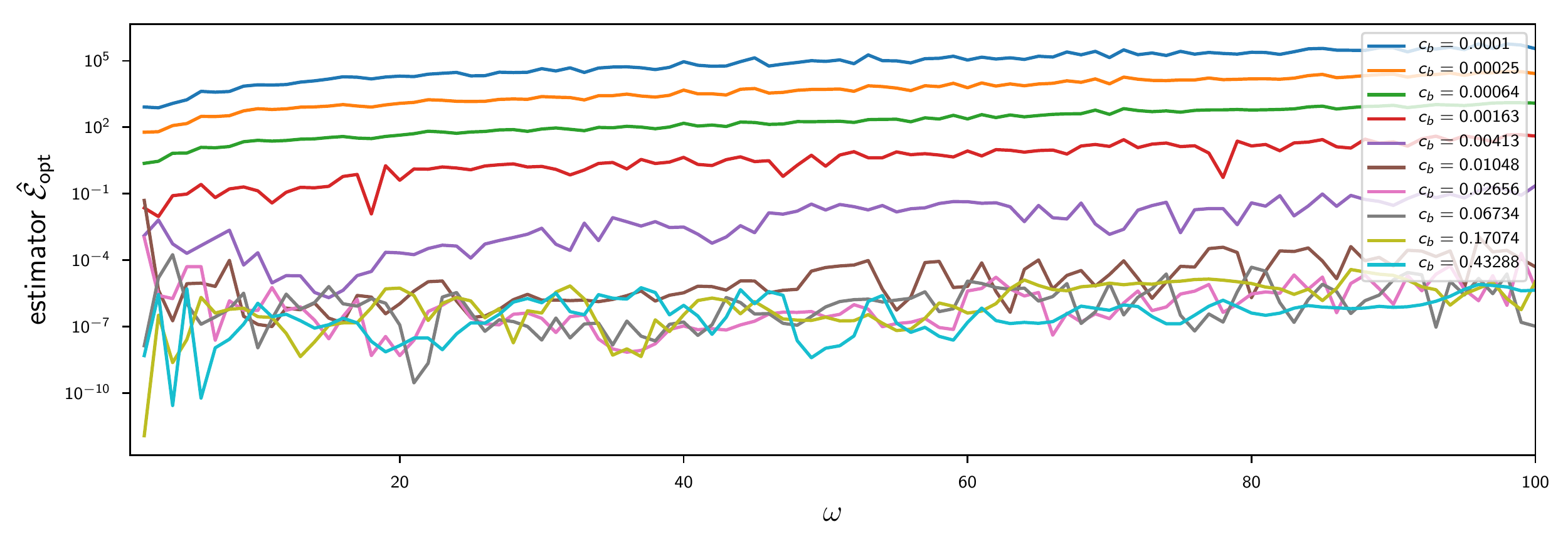}
}
\caption{Estimated amplitudes $M$ for the resummed estimator defined in Eq.~\eqref{eq:eopt_2d} for WMAP, as a function of frequency. The different curves show $10$ log-spaced $c_b$ values between $c_b=0.0001$ and $c_b=1.0$. The estimator for each $\omega$ has been maximized over $\phi$. The Gaussian mean was subtracted and the Fisher matrix estimated from Monte Carlo. The amplitudes give an impression of the theoretical constraints we obtain, as none of the values is significant within the ensemble of Monte Carlo maps.} 
\label{fig:resEXP}
\end{figure}

Our results for WMAP are shown in Fig.~\ref{fig:resEXP}. We show results for 10 example values of $c_b$, from our set of 100 logarithmically spaced sampling points between $c_b=0.0001$ and $c_b=1.0$. The $c_b$ values have been picked within the interesting range found by examining the Fisher matrix contributions of different N as illustrated in Fig.~\ref{fig:res4}. 
The estimated amplitudes $M$ in Fig.~\ref{fig:resEXP} span a wide range of magnitudes. Numbers above one would mean a dense production of particles, with a low aplitude $c_b$, so that individual profiles do not stand out of the noise. On the high $c_b$ end we estimated amplitudes as low as $10^{-9}$. Such values should not be taken at face value but are an artifact of our mathematical description in terms of correlation functions, which does not take into account the discrete number of events (i.e. an experiment might see exactly zero events). As descussed in the beginning of Sec. \ref{sec:cmb_estimator}, we cannot constrain $M$ below $\sim \ell_{max}^{-2} \sim 10^{-6}$, as this is the number of available pixels. In addition, for such high values of $c_b$, one would see the profile in a pixel histogram or even by eye. By comparing at what $c_b$ we get unphysical constraints ($c_b\sim 0.004$) with the Fisher matrix contributions at this $c_b$ in Fig.~\ref{fig:res4}, we see that the largest $N$-point functions of interest are around $N\simeq 40$ in this model.

To assess statistical significance we again define a peak statistic 
\be
\label{eq:peakestimator2}
\mathcal{E}_{\rm peak}= \max_{\omega_i,\phi_j} |\mathcal{E}_{\rm opt}^{\omega_i,\phi_i}|.
\ee
In addition to that, we also  use a simple integrated statistic, where we sum over the amplitudes of of all frequencies and phases, i.e.
\be
\label{eq:sumestimator2}
\mathcal{E}_{\rm sum} = \sum_{\omega_i,\phi_j} |\mathcal{E}_{\rm opt}^{\omega_i,\phi_i}|
\ee
This statistic can be interpreted as marginalizing over the parameter space spanned by the frequency $\omega$ and phase $\phi$ as can be seen from the likelihood interpretation in Sec. \ref{sec:profilefinding}.

\begin{figure}
\resizebox{0.8\hsize}{!}{
\includegraphics{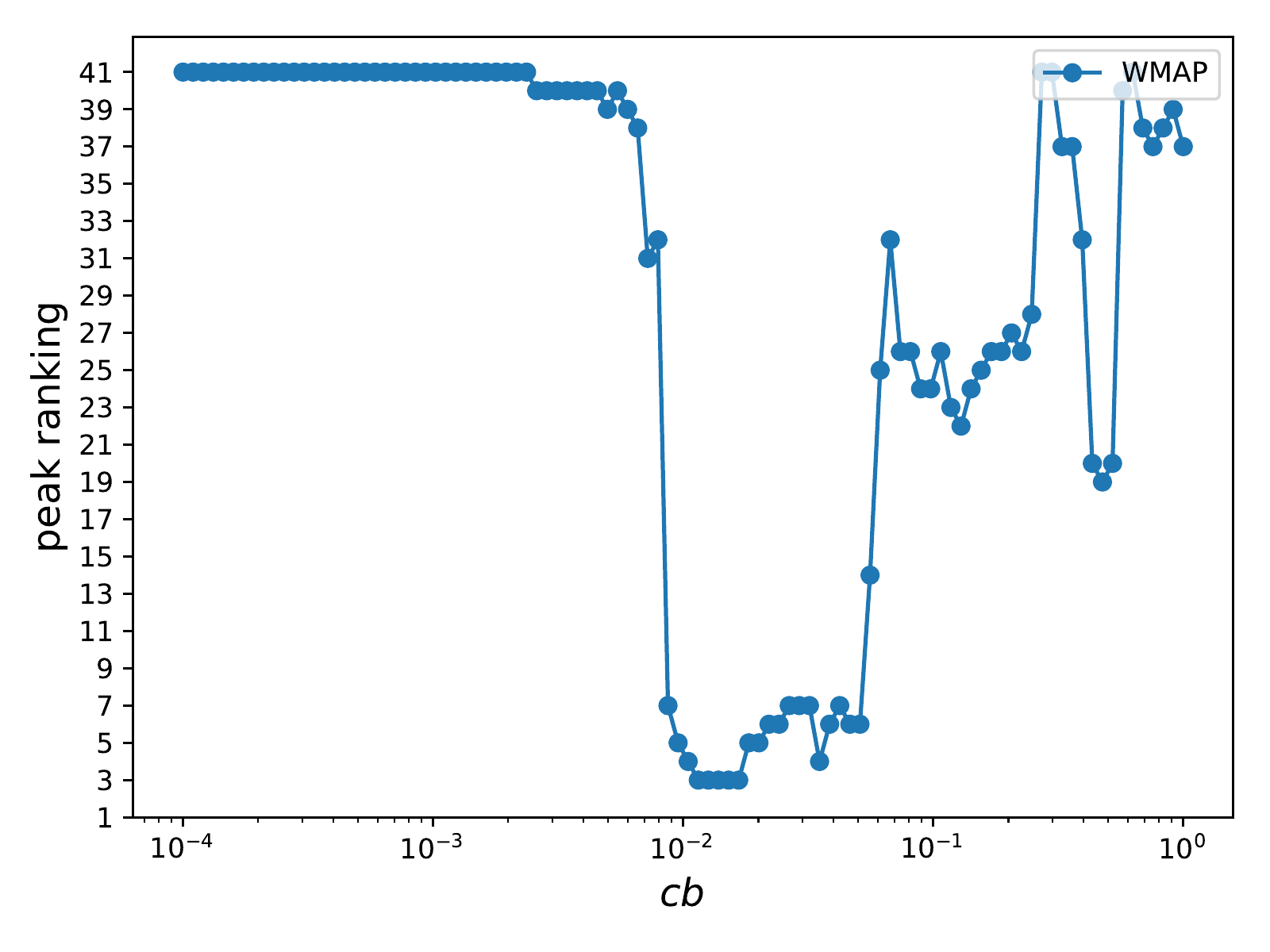}\includegraphics{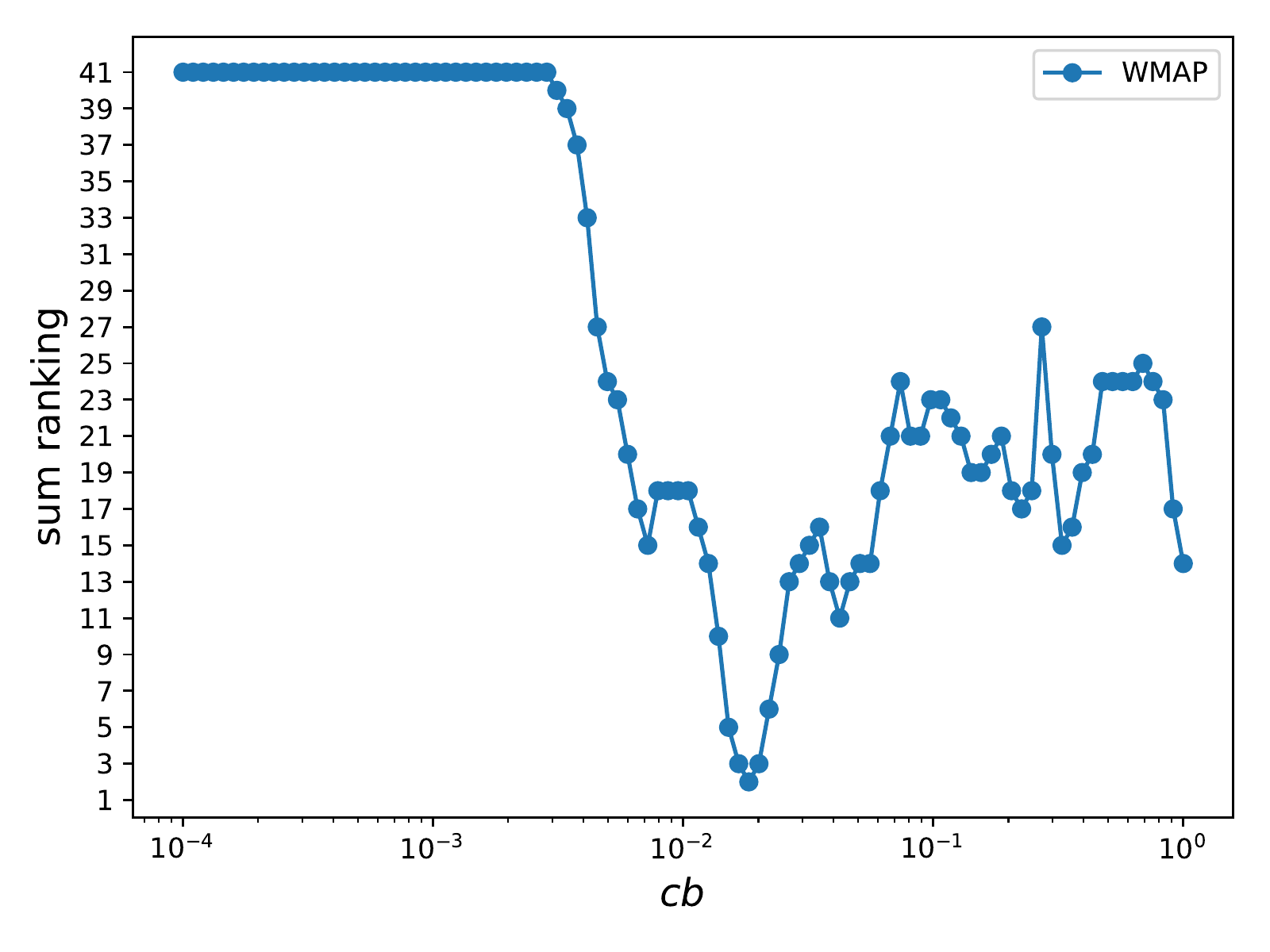}
}
\caption{Top: Ranking of highest peak (left) and the sum over all estimators (right) in the WMAP data, compared to 40 Monte Carlo maps, as a function of $c_b$, using the resummed estimator defined in Eq.~\eqref{eq:eopt_2d}. Here rank 1 means the highest peak in the ensemble of maps and rank 41 the lowest. We find that for low $c_b$ the data is the least significant in our ensemble. At its highest peak the data ranks second. Overall we do not find a significant excess in the data.}
\label{fig:res5}
\end{figure}

The results for the peak and sum statistics are shown in Fig.~\ref{fig:res5}. At its highest peak, the data ranks second among the ensemble of data and simulation maps. This is a moderate peak, especially considering the scan over frequencies which adds a look elsewhere penalty, so we remain consistent with Gaussianity. Interestingly for low $c_b$ the data is actually the least significant (by a small difference). This low range of $c_b$ is dominated by the lowest $N$ and is likely simply a statistical fluctuation. We have not attempted to derive precise constraints on primordial parameters from our results, because of the unknown estimator PDF.

\section{Conclusion}
\label{sec:conclusion}

The present study, to our knowledge, provides the first observational exploration of primordial non-Gaussianity at higher N-point functions than the trispectrum. The specific model under consideration is that of periodic heavy particle production from oscillating masses in \cite{Flauger:2016idt}. We have examined the arbitrary-N equivalent of the well known KSW estimator \cite{Komatsu:2003iq} for separable bispectra. We have shown how to calculate the estimator expectation value and covariance in the specific case of a Poisson type hierarchy of N-point functions, and defined an optimal resummed estimator that takes into account this covariance. We show that this estimator is equivalent to a profile finding likelihood. 

We have applied our estimators to CMB data from the WMAP satellite. Our results are consistent with Gaussianity, probing a new form of non-Gaussianity that was not previously studied, and is likely orthogonal to previous shapes due to the oscillation in the density perturbations induced by the particles. Our analysis illustrates properties of high-$N$ non-Gaussianity search, that are important beyond the present model, in particular the non-Gaussianity of the estimator PDF, and how to deal with correlations between different $N$. 

Our analysis is only the beginning of higher N-point function searches. In particular we have only examined the Poisson limit of the shape functions, excluding interactions between the produced particles, and we have only probed a specific form of mass function. Momentum space estimators for $N$-point functions would become a powerful tool once interactions between particles are taken into account. Further theoretical and observational studies are needed to better understand what constraints on primordial physics can be obtained from this kind of analysis and realize the full potential of the data.

\section*{Acknowledgments}

We thank Mehrdad Mirbabayi, Leonardo Senatore and Eva Silverstein for extensive discussions, Leonardo Senatore in particular for proposing the $N$-point estimator, Mehrdad Mirbabayi in particular for deriving the real space profile in appendix B, and Eva Silverstein in particular for proposing the saddle point approximation in appendix C. MM thanks the hospitality of Stanford Institute for Theoretical Physics where part of this work was completed. Research at Perimeter Institute is supported by the Government of Canada
through Industry Canada and by the Province of Ontario through the Ministry of Research \& Innovation.
Some computations were performed on the GPC cluster at the SciNet HPC Consortium.
SciNet is funded by the Canada Foundation for Innovation under the auspices of Compute Canada,
the Government of Ontario, and the University of Toronto.
KMS was supported by an NSERC Discovery Grant and an Ontario Early Researcher Award.

\bibliography{particle_ng_analysis}

\appendix

\section{Estimator properties and optimal resumming}
\label{app:resum}

In this appendix, we derive Eqs.~(\ref{eq:eopt_3d}),~(\ref{eq:eopt_2d}) for the resummed estimator $\hE_{\rm opt}$,
in the 3-d and 2-d cases respectively.

\subsection{3-D case}

First we recall the setup from~\S\ref{sec:estimator3d}.
We consider an inflationary model which generates the following $N$-point correlation function:
\be
\big\langle \zeta_{\k_1} \cdots \zeta_{\k_N} \big\rangle_c
  = \alpha \left( \prod_{i=1}^N h(k_i) \right) (2\pi)^3 \delta^3\Big(\sum\k_i\Big)
    \hspace{1cm} \mbox{for $N\ge 3$} \label{eq:zetaN_3d}
\ee
For $N=2$, we assume that the power spectrum gets the following $\bigoh(\alpha)$ contribution: 
\be
\big\langle \zeta_{\k} \zeta_{\k'}^* \big\rangle = \Big( P(k) + \alpha h(k)^2 \Big) \, (2\pi)^3 \delta^3(\k-\k')  \label{eq:zeta2_3d}
\ee
As an ansatz, we will assume that the optimal estimator $\hE_{\rm opt}$ for $\alpha$ is a linear
combination of the $N$-point estimators $\hE_N$ defined by:
\be
\hE_N = \frac{1}{N!} \int d^3\x \, \psi(\x)^N  \hspace{1.5cm} \mbox{where } \psi_{\k} = \frac{h(k)}{P(k)} \zeta_\k
\ee
We will use this definition for all $N\ge 0$, but note that $\hE_0 = V$ and $\hE_1 = 0$,
where $V$ is the 3-d volume (assumed finite).

Define $\phi(r)$ to be the correlation function of the $\psi$-field:
\be
\phi(r) = \big\langle \psi(\x) \psi(\x+\r) \big\rangle = \int \frac{d^3\k}{(2\pi)^3} \frac{h(k)^2}{P(k)} e^{i\k\cdot\r}
\ee
For use later in this appendix, we calculate the connected $N$-point function $\langle \psi(\x)^N \rangle_c$.
For $N \ge 3$, we have:
\ba
\big\langle \psi(\x)^N \big\rangle_c
&=& \int \frac{d^3\k_1}{(2\pi)^3} \cdots \frac{d^3\k_N}{(2\pi)^3} \, \left( \prod_{i=1}^N \frac{h(k_i)}{P(k_i)} \right) e^{i(\sum\k_i)\cdot\x} \,
  \big\langle \zeta_{\k_1} \cdots \zeta_{\k_N} \rangle_c \nn \\
&=& \alpha \int \frac{d^3\k_1}{(2\pi)^3} \cdots \frac{d^3\k_N}{(2\pi)^3} \, \left( \prod_{i=1}^N \frac{h(k_i)^2}{P(k_i)} \right) (2\pi)^3 \delta^3\Big(\sum\k_i\Big) \nn \\
&=& \alpha \int d^3\r \, \prod_{i=1}^n \left( \int \frac{d^3\k_i}{(2\pi)^3} \, \frac{h(k_i)^2}{P(k)} e^{i\k\cdot\r} \right) \nn \\
&=& \alpha \int d^3\r \, \phi(r)^N \hspace{1.5cm} \mbox{for } N \ge 3  \label{eq:psiN_3d}
\ea
To get the third line, we have used the identity $(2\pi)^3 \delta^3(\sum\k_i) = \int d^3\r \, e^{i(\sum\k_i)\cdot\r}$.
For $N=2$, a short calculation using Eq.~(\ref{eq:zeta2_3d}) gives:
\be
\big\langle \psi(\x)^2 \big\rangle_c = \phi(0) + \alpha \int d^3\r \, \phi(r)^2  \label{eq:psi2_3d}
\ee
It will be convenient to introduce the quantity $F_N$, defined by:
\be
F_N = \frac{V}{N!} \int d^3\r \, \phi(r)^N  \hspace{1cm} \mbox{for } N \ge 1
\ee
We will use this definition for all $N\ge 1$, but we note that $F_1=0$ since $\int d^3\r \, \phi(\r) = 0$.
As an aside, $F_N$ is the Fisher matrix element for the $N$-point signal in this model,
but we will not need this interpretation.  Then Eqs.~(\ref{eq:psiN_3d}),~(\ref{eq:psi2_3d})
can be written:
\be
\big\langle \psi(\x)^N \big\rangle_c = \phi(0) \delta_{N,2} + \alpha \frac{N!}{V} F_N   \hspace{1.5cm} \mbox{for } N \ge 2
\ee
We next compute the expectation value $\langle \hE_N \rangle$ to first order in $\alpha$,
including disconnected contributions of the form $\langle \psi^2 \rangle^{N-2M} \, \langle \psi^M \rangle_c$.
To organize all contributions, it is convenient to use a generating function formalism.
We define the generating function $\hE(t)$ by:
\ba
\hE(t) = \sum_{N\ge 0} t^N \hE_N = \int d^3\x \, e^{t \psi(\x)}
\ea
and compute $\langle \hE(t) \rangle$ as follows:
\ba
\langle \hE(t) \rangle
&=& \int d^3\x \, \big\langle e^{t \psi(\x)} \big\rangle \nn \\
&=& \int d^3\x \, \exp\bigg( \sum_{N\ge 2} \frac{t^N}{N!} \langle \psi(\x)^N \rangle_c \bigg) \nn \\
&=& \int d^3\x \, \exp\bigg( \frac{t^2}{2} \phi(0) + \frac{\alpha}{V} \sum_{N\ge 2} t^N F_N \bigg) \nn \\
&=& e^{t^2 \phi(0)/2} \bigg( V + \alpha \sum_{N\ge 2} t^N F_N \bigg) + \bigoh(\alpha^2)  \label{eq:Ebar_3d}
\ea
To get the second line, we used the cumulant expansion theorem to write $\langle e^{t\psi(\x)} \rangle$
in terms of {\em connected} cumulants $\langle \psi(\x)^N \rangle_c$.
The generating function in Eq.~(\ref{eq:Ebar_3d}) encodes the expectation values $\langle \hE_N \rangle$,
via the series expansion $\langle \hE(t) \rangle = \sum_N t^N \langle \hE_N \rangle$.

Next we compute the expectation value $\langle \hE_M \hE_N \rangle$, to zeroth order in $\alpha$
(i.e.~assuming Gaussian statistics).
Again it is convenient to use a generating function formalism, and compute $\langle \hE(t) \hE(t') \rangle$ as follows:
\ba
\big\langle \hE(t) \hE(t') \big\rangle
&=& \int d^3\x \, d^3\x' \, \big\langle e^{t\psi(\x) + t'\psi(\x')} \big\rangle \nn \\
&=& \int d^3\x \, d^3\x' \, \exp\left[ \frac{1}{2} \Big\langle \big( t\psi(\x) + t'\psi(\x') \big)^2 \Big\rangle \right] \nn \\
&=& \int d^3\x \, d^3\x' \, \exp\left[ \frac{1}{2} \Big( t^2 \phi(0) + 2 tt' \phi(\x-\x') + t'{}^2 \phi(0) \Big) \right] \nn \\
&=& e^{t^2 \phi(0)/2} e^{t'{}^2 \phi(0)/2} \int d^3\x \, d^3\x' \, \bigg(\sum_{N\ge 0} \frac{(tt')^N}{N!} \phi(\x-\x')^N \bigg) \nn \\
&=& e^{t^2 \phi(0)/2} e^{t'{}^2 \phi(0)/2} \bigg(V^2 + \sum_{N\ge 2} (tt')^N F_N \bigg)  \label{eq:E2bar_3d}
\ea
For reasons that will be apparent shortly, we define new estimators $\hE'_N$ by taking the
generating function $\hE'(t) = \sum_N t^N \hE'_N$ to be:
\ba
\hE'(t) = e^{-t^2\phi(0)/2} \hE(t)  \label{eq:Eprime_def_3d}
\ea
Under the change of variable $\hE \rightarrow \hE'$, Eqs.~(\ref{eq:Ebar_3d}) and~(\ref{eq:E2bar_3d}) simplify as follows:
\be
\big\langle \hE'(t) \big\rangle = V + \alpha \sum_{N \ge 2} t^N F_N
\hspace{1.5cm}
\big\langle \hE'(t) \hE'(t') \big\rangle = V^2 + \sum_{N\ge 2} (tt')^N F_N   \label{eq:Eprime_bar_3d}
\ee
Equivalently, unpacking the generating functions, the mean and covariance of $\hE'_N$ have the following simple forms:
\begin{align}
\big\langle \hE'_N \big\rangle &= \alpha F_N & (N\ge 2) \\
\Cov(\hE'_M, \hE'_N) &= F_N \delta_{MN} & (M,N\ge 2)
\end{align}
We have introduced $\hE'_N$ since it results in these algebraic simplifications.
Another motivation for $\hE'_N$ is that it is actually the optimal estimator for the $N$-point
function in Eq.~(\ref{eq:zetaN_3d}).  Generally speaking, the optimal estimator for an $N$-point function
contains lower-order terms of orders $(N-2), (N-4), \cdots$.  In our case, including these terms gives
estimator $\hE'_N = \hE_N - (\langle \psi^2 \rangle/2) \hE_{N-2} + (\langle \psi^2 \rangle^2/8) \hE_{N-4} + \cdots$.
This is precisely the relation between $\hE'_N$ and $\hE_N$ encoded by the generating function
identity~(\ref{eq:Eprime_def_3d}).

Our goal is to derive the optimal estimator, which is a linear combination (over $N$) of estimators $\hE'_N$:
\be
\hE_{\rm opt} = \sum_{N\ge 2} W_N \hE'_N  \label{eq:Eopt_W_3d}
\ee
with weights $W_N$ which minimize the variance $\mbox{Var}(\hE_{\rm opt})$, subject to the constraint
$\langle \hE_{\rm opt} \rangle = \alpha$.  By Eqs.~(\ref{eq:Eprime_bar_3d}), the mean and variance of
$\hE_{\rm opt}$ are:
\be
\langle \hE_{\rm opt} \rangle = \alpha \sum_{N\ge 2} W_N F_N  \hspace{1.5cm}  \Var(\hE_{\rm opt}) = \sum_{N\ge 2} W_N^2 F_N  \label{eq:Eopt_mean_var_3d}
\ee
To solve the constrained minimization, we introduce a Lagrange multiplier $\lambda$,
and differentiate both parts of Eq.~(\ref{eq:Eopt_mean_var_3d}) with respect to $W_N$, obtaining:
\be
W_N F_N = \lambda F_N
\ee
which implies $W_N = \lambda$ (independent of $N$).
The value of $\lambda$ is determined by the constraint $\langle \hE \rangle = \alpha$.
Using Eq.~(\ref{eq:Eopt_mean_var_3d}), we get $\lambda = F_{\rm tot}^{-1}$, where:
\ba
F_{\rm tot}
&=& \sum_{N \ge 2} F_N \nn \\
&=& \sum_{N \ge 2} \frac{V}{N!} \int d^3\r \, \phi(r)^N \nn \\
&=& \int d^3\r \, \Big( e^{\phi(r)} - 1 \Big)  \label{eq:Ftot_3d}
\ee
where we have used $\int d^3\r \, \phi(r) = 0$ in the last step.
We get our final expression for $\hE_{\rm opt}$ by plugging $W_N = F_{\rm tot}^{-1}$ into Eq.~(\ref{eq:Eopt_W_3d}) and simplifying as follows:
\ba
\hE_{\rm opt}
&=& F_{\rm tot}^{-1} \sum_{N\ge 2} \hE'_N \nn \\
&=& F_{\rm tot}^{-1} \Big( \hE'(1) - \hE'_0 - \hE'_1 \Big) \nn \\
&=& F_{\rm tot}^{-1} \Big( e^{-\phi(0)/2} \hE(1) - \hE_0 - \hE_1 \Big) \nn \\
&=& F_{\rm tot}^{-1} \int d^3\x \, \Big( e^{-\phi(0)/2} e^{\psi(\x)} - 1 \Big)
\ea
In the second line, we have used the identity $\hE'(t) = \sum_N t^N \hE'_N$ evaluated at $t=1$.
In the third line, we have changed variable $\hE' \rightarrow \hE$ using Eq.~(\ref{eq:Eprime_def_3d}).
The final result for $\hE_{\rm opt}$ agrees with Eq.~(\ref{eq:eopt_3d}) in the main text, after changing notation $\phi(0) \rightarrow \langle \psi^2 \rangle$.

\subsection{2-D case}

Now we analyze the 2-d CMB case from~\S\ref{sec:cmb_estimator}.
Our goal is to derive the expression for $\hE_{\rm opt}$ in Eq.~(\ref{eq:eopt_2d}).

First we recall the setup in the 2-d CMB case.
We do not assume that the CMB noise is isotropic, and our calculation allows an arbitrary noise covariance matrix $N$.
The inverse signal + noise covariance $C^{-1} = (S+N)^{-1}$ appears frequently, and we represent it by the tensor
$C^{-1}_{l_1m_1,l_2m_2}$, defined by $(C^{-1}a)_{lm} = C^{-1}_{lm,l'm'} a_{l'm'}$.
The two-point function of $C^{-1}a$ is then given by:
\be
\big\langle (C^{-1}a)_{lm} (C^{-1}a)_{l'm'}^* \big\rangle = C^{-1}_{lm,l'm'}  \label{eq:cinva_2pt}
\ee
We are interested in the following $N$-point signal:
\be
\big\langle a_{l_1m_1} \cdots a_{l_Nm_N} \big\rangle_c = \alpha \sum_n f_n \int d^3\r \, \left( \prod_{i=1}^N M_{l_i}^n(r) Y_{l_im_i}^*(\r) \right)
 \hspace{1.5cm} \mbox{for $N\ge 3$} \label{eq:almN_2d}
\ee
where we have introduced the prefactor $\alpha$ to keep track of the overall power of non-Gaussianity in the calculations which follow.
For $N=2$, we write the two-point function as:
\be
\big\langle a_{l_1m_1} a_{l_2m_2}^* \big\rangle = C_{l_1m_1,l_2m_2} + \alpha \sum_n f_n \int d^3\r \, M_{l_1}^n(r) M_{l_2}^n(r) \, Y_{l_1m_1}^*(\r) Y_{l_2m_2}(\r)
\ee
The 2-d CMB case is more complicated than the 3-d primordial case: the noise model is anisotropic, and the
$N$-point signal in Eq.~(\ref{eq:almN_2d}) is more complicated algebraically than Eq.~(\ref{eq:zetaN_3d}).
However, these complications will not affect the calculations much, and so we will present the derivation
in streamlined form, since it is similar to the 3-d case from the previous section.

We define:
\be
\hE^n_N(\r) = \frac{1}{N!} \psi_n(\r)^N  \hspace{1.5cm} \mbox{where } \psi_n(\r) = \sum_{lm} M_l^n(r) (C^{-1}a)_{lm} Y_{lm}(\hr)
\ee
Note that in the 2-d case, there is no integral in the definition of $\hE$.  The Gaussian two-point function of $\psi$ is:
\ba
\phi_{nn'}(\r,\r')
&=& \big\langle \psi_n(\r) \psi_{n'}(\r') \big\rangle \nn \\
&=& \sum_{lml'm'} M_l^n(r) M_{l'}^{n'}(r') C^{-1}_{lm,l'm'} Y_{lm}(\hr) Y_{l'm'}^*(\hr') 
\ea
using Eq.~(\ref{eq:cinva_2pt}) to get the second line.
We calculate the expectation value $\langle \psi^N \rangle_c$ as follows:
\ba
\big\langle \psi_n(\r)^N \big\rangle_c
&=& \sum_{l_im_il'_im'_i} \left( \prod_{i=1}^N M_{l_i}^n(r) C^{-1}_{l_im_i,l'_im'_i} Y_{l_im_i}(\hr) \right) \, \big\langle a_{l'_1m'_1} \cdots a_{l'_Nm'_N} \big\rangle_c \nn \\
&=& \alpha \sum_{n'} f_{n'} \int d^3\r' \sum_{l_im_il'_im'_i} \left( \prod_{i=1}^N M_{l_i}^n(r) M_{l'_i}^{n'}(r') C^{-1}_{l_im_i,l'_im'_i} Y_{l_im_i}(\hr) Y_{l'_im'_i}^*(\hr') \right) \nn \\
&=& \alpha \sum_{n'} f_{n'} \int d^3\r' \phi_{nn'}(\r,\r')^N \hspace{1cm} \mbox{for $N\ge 3$}  \label{eq:psiN_2d}
\ea
For $N=2$ we have:
\be
\big\langle \psi_n(\r)^2 \big\rangle_c = \phi_{nn}(\r,\r) + \alpha \sum_{n'} f_{n'} \int d^3\r' \phi_{nn'}(\r,\r')^2  \label{eq:psi2_2d}
\ee
Define the generating function $\hE_n(t,\r)$ by:
\ba
\hE_n(t,\r) = \sum_N t^N \hE^n_N(\r) = e^{t\psi_n(\r)}
\ea
We calculate the mean $\langle \hE \rangle$ in generating function form, to first order in $\alpha$:
\ba
\big\langle \hE_n(t,\r) \big\rangle
&=& \big\langle e^{t\psi_n(\r)} \big\rangle \nn \\
&=& \exp\bigg( \sum_{N\ge 1} \frac{t^N}{N!} \big\langle \psi_n(r\n)^N \big\rangle_c \bigg) \nn \\
% &=& \exp\bigg( \frac{t^2}{2} \phi_{nn}(\r,\r) + \alpha \sum_{N\ge 2} \frac{t^N}{N!} \sum_{n'} f_{n'} \int d^3\r' \, \phi_{nn'}(\r,\r')^N \bigg) \nn \\
&=& e^{t^2\phi_{nn}(\r,\r)/2} \bigg( 1 + \alpha \sum_{N\ge 2} \frac{t^N}{N!} \sum_{n'} f_{n'} \int d^3\r' \, \phi_{nn'}(\r,\r')^N \bigg) + \bigoh(\alpha^2)
\ea
using Eqs.~(\ref{eq:psiN_2d}),~(\ref{eq:psi2_2d}) to get the last line.
We calculate the two-point expectation value $\langle \hE \hE \rangle$ in generating function form, to zeroth order in $\alpha$:
\ba
\big\langle \hE_n(t,\r) \, \hE_{n'}(t',\r') \big\rangle
&=& \big\langle e^{t\psi_n(\r) + t'\psi_{n'}(\r')} \big\rangle \nn \\
&=& \exp \bigg( \frac{1}{2} \Big\langle \big( t\psi_n(\r) + t'\psi_{n'}(\r') \big)^2 \Big\rangle \bigg) \nn \\
&=& \exp \bigg( \frac{1}{2} \big( t^2 \phi_{nn}(\r,\r) + 2tt' \phi_{nn'}(\r,\r') + t'{}^2 \phi_{n'n'}(\r',\r') \big) \bigg) \nn \\
&=& e^{t^2\phi_{nn}(\r,\r)/2} e^{t'{}^2\phi_{n'n'}(\r',\r')/2} \bigg( \sum_{N\ge 0} \frac{(tt')^N}{N!} \phi_{nn'}(\r,\r')^N \bigg)
\ea
Now define alternate estimators $\hE^{'n}_N(\r)$ by taking the generating function $\hE'_n(t,\r) = \sum_N t^N \hE^{'n}_N(\r)$ to be:
\be
\hE'_n(t,\r) = e^{-t^2\phi_{nn}(\r)/2} \hE_n(t,\r) = e^{-t^2\phi_{nn}(\r)/2} e^{t\psi_n(\r)} \label{eq:Eprime_def_2d}
\ee
Unpacking generating functions, the mean and covariance of $\hE^{'n}_N(\r)$ are:
\ba
\big\langle \hE^{'n}_N(\r) \big\rangle &=& \frac{\alpha}{N!} \sum_{n'} f_{n'} \int d^3\r' \, \phi_{nn'}(\r,\r')^N  \hspace{1cm} (\mbox{for } N\ge 2) \label{eq:Eprime_mean_2d} \\
\mbox{Cov}\big(\hE^{'m}_M(\r), \hE^{'n}_N(\r') \big) &=& \frac{1}{N!} \phi_{nn'}(\r,\r')^N \delta_{MN} \label{eq:Eprime_cov_2d}
\ea
Our goal is to derive the optimal estimator, which is a linear combination (over $N,n,\r$) of estimators $\hE^{'n}_N(\r)$:
\be
\hE_{\rm opt} = \sum_{N\ge 2} \sum_n \int d^3\r \, W_N^n(\r) \hE^{'n}_N(\r)  \label{eq:Eopt_def_2d}
\ee
with weight function $W_N^n(\r)$ which minimizes $\Var(\hE_{\rm opt})$ subject to the constraint $\langle \hE_{\rm opt} \rangle = \alpha$.
By Eqs.~(\ref{eq:Eprime_mean_2d}),~(\ref{eq:Eprime_cov_2d}), the mean and variance of $\hE_{\rm opt}$ are:
\ba
\big\langle \hE_{\rm opt} \big\rangle &=& \alpha \sum_{N\ge 2} \frac{1}{N!} \sum_{nn'} \int d^3\r \, d^3\r' \, W_N^n(\r) \phi_{nn'}(\r,\r')^N f_{n'} \label{eq:Eopt_mean_2d} \\
\mbox{Var}(\hE_{\rm opt}) &=& \sum_{N\ge 2} \frac{1}{N!} \sum_{nn'} \int d^3\r \, d^3\r' \, W_N^n(\r) W_N^{n'}(\r') \phi_{nn'}(\r,\r')^N \label{eq:Eopt_var_2d}
\ea
Introducing a Lagrange multiplier $\lambda$, the optimal weights satisfy the equation:
\be
\frac{1}{N!} \sum_{n'} \int d^3\r' \, \phi_{nn'}(\r,\r')^N W_N^{n'}(\r') = \frac{\lambda}{N!} \sum_{n'} \int d^3\r' \, \phi_{nn'}(\r,r')^N f_{n'}
\ee
which has solution:
\be
W_N^{n'}(\r') = \lambda f_{n'} \hspace{1cm} \mbox{(independent of $N,\r$)}  \label{eq:Wopt_2d}
\ee
The value of $\lambda$ is determined by the constraint $\langle \hE_{\rm opt} \rangle = \alpha$.
By Eq.~(\ref{eq:Eopt_mean_2d}), we get $\lambda = F^{-1}$, where
\ba
F &=& \sum_{N\ge 2} \frac{1}{N!} \sum_{nn'} \int d^3\r \, d^3\r' \, f_n f_{n'} \phi_{nn'}(\r,\r')^N \nn \\
&=& \sum_{nn'} \int d^3\r \, d^3\r' \, f_n f_{n'} \Big( \exp\big(\phi_{nn'}(\r,\r')\big) - 1 \Big) \label{eq:F_2d}
\ea
By Eq.~(\ref{eq:Eopt_var_2d}), the quantity $F$ is related to the variance of $\hE_{\rm opt}$ by:
\be
\Var(\hE_{\rm opt}) = F^{-1}
\ee
We get our final expression for $\hE_{\rm opt}$ by plugging $W^n_N(\r) = F^{-1} f_n$ into Eq.~(\ref{eq:Eopt_def_2d}):
\ba
\hE_{\rm opt}
&=& F^{-1} \sum_n f_n \int d^3\r \, \sum_{N\ge 2} \hE^{'n}_N(\r) \nn \\
&=& F^{-1} \sum_n f_n \int d^3\r \, \Big( \hE'_n(\r) - \hE^{'n}_0 - \hE^{'n}_1 \Big) \nn \\
&=& F^{-1} \sum_n f_n \int d^3\r \, \Big( e^{-\phi_{nn}(\r)/2} e^{\psi_n(\r)} - 1 \Big)
\ea
where we have used $\int d^3\r \, \hE^{'n}_1(\r) = 0$ in the last step.
The final result for $\hE_{\rm opt}$ agrees with Eq.~(\ref{eq:eopt_2d}) in the main text, after changing notation $\phi_{nn}(\r) \rightarrow \langle \psi_n(\r)^2 \rangle$.

\section{Review of the classical Poisson particle production process}
\label{sec:poisson}
While the non-Gaussian shape under consideration in this paper is part of a complete microphysical model, its features can be understood physically as classical particle production. The resulting shape function was derived in ~\cite{Mirbabayi:2014jqa}. We briefly review this derivation in momentum space, generalised to arbitrary mass functions. We also discuss the shape in real-space to build intuition about its observability.

\subsection{Particle production in momentum space}

In the Poisson model we produce particles independently from each other at times $\eta_n$ and positions $x_i$. In addition, the mass function of the particles is time dependent $m(t)$, which leads to a continuous emission of curvature perturbations of each particle after its production.

For a single particle of mass $m(t)$ produced at $\x = 0$ and $\eta=\eta_n$, the induced curvature perturbations is
\be\label{zr}
\zeta_\k(\eta = 0)= \frac{1}{2\ep\mpl^2}\int_{\eta_n}^0\frac{d\eta}{\eta}\dot m \frac{g(k\eta)}{k^3},
\ee
where $g(k\eta)$ is the de Sitter retarded Greens function in the limit $k \eta \rightarrow 0$, over-dot is $d/dt$ and where $\ep = - \dot H/H^2$. The curvature perturbation in momentum space from a random distribution of particles is thus
\be\label{zk}
\zeta(\k) = \sum_{n} \frac{h(k\eta_n)}{k^3} \sum_{i} X_{n,i} e^{i\k\cdot x_i} 
\ee
where 
\be
h(k\eta_n) = \frac{1}{2 \ep \mpl^2} \int_{\eta_n}^0 \frac{d\eta}{\eta} g(k\eta) \dot m(t).
\ee
Here $X_{n,i}$ are random variables ($X \in [0,1]$) indexing when and where events are produced. The connected part of the correlation function of the production events is 
\be \label{poisson}
\expect{X_{n,i} \ X_{m,j} \ X_{l,k} \dots }_c  = \bar{n} a^3 \delta v_i \ \delta_{ijk \dots}  \delta_{nml \dots} 
\ee
where $\bar{n}$ is the average proper density of particles and $\delta v_i$ is the small comoving volume associated with the random variable at position $x_i$. The fully connected part of the correlation function is therefore induced by each single particle independently. Using Eq. \eqref{zk} and Eq. \eqref{poisson} one finds
\be
\expect{\zeta(\k_1) \zeta(\k_2) \zeta(\k_3) \dots}'_c = \bar{n} \sum_{n} (H \eta_n)^{-3} \prod_{i} \frac{h(k_i\eta_n)}{k_i^3}.  
\ee

\subsection{Particle production in real space}
\label{sec:realspace}

To build intuition about the observability of a single particle produced at $\eta_n$, we derive the real space profile from the momentum space perturbation in Eq. \eqref{zr}. To find the real space profile, we first need the emission of a delta-function source
\be
f(\eta,r) = \int \frac{d^3 \k}{(2\pi)^3} e^{-i\k\cdot \r}\frac{g(k\eta)}{k^3}.
\ee
Using $g(k\eta) = -k^2 \d_k [\sin(k\eta)/k]$ we obtain
\be
f(\eta,r) = \frac{1}{(2\pi)^2}\int_{-\infty}^{\infty}\frac{dk}{k+i\ep}\sin k\eta \ \cos kr
\ee
where we used the symmetry of the integrand and added $-i\pi \delta(k)\sin k\eta \ \cos kr \ (=0)$ to shift the pole in the complex plane. Writing sine and cosine in terms of exponentials we can perform the contour integral to get
\be
f(\eta,r)= \frac{1}{8\pi}[\theta(-\eta-r)+\theta(r-\eta)-\theta(\eta-r)-\theta(\eta+r)]=\frac{1}{4\pi}\theta(-\eta-r).
\ee
Substituting in \eqref{zr} we get
\be
\zeta(0,\r) = \frac{1}{8\pi \ep\mpl^2}\int_{\eta_n}^r \frac{d\eta}{\eta}\dot m .
\ee
Now we can change the variable of integration to $t$ and get $\zeta(0,\r)$ as a function of the difference between the initial mass and $m(t_r)$ where $t_r$ is the time when the past light-cone of the point $(\eta=0,\r)$ crosses the particle world-line:
\be
t_r = -H^{-1}\log(Hr).
\ee
Note that the profile has finite extent up to $r = -\eta_n$ and a definite sign.
Therefore, we get
\be
\zeta(0,\r) = -\frac{H}{8\pi \ep\mpl^2} [m(t_r-t_n)-m(t_n)].
\ee
For example, in a model with $m^2(t) = \mu^2 + m_0^2 \sin^2 \omega t$, the production events are at $t_n$ when the sine vanishes, so that
\be
\zeta(0,\r) \sim -\frac{H}{8\pi \ep\mpl^2} \left[\sqrt{\mu^2 + m_0^2 \sin^2 (\frac{\omega}{H}\log Hr)}-\mu\right].
\ee
 
In our explicit microphysical model with
\be
m^2 = \mu^2 + 2 g^2 f^2 \cos \phi/f.
\ee
we can relate the amplitude of the profile to the model parameters as follows. In the regime $\mu\gg gf$, we have
\be
\zeta_{\rm max} \sim \frac{H}{8\pi\ep \mpl^2} [\sqrt{\mu^2 + 2g^2 f^2} -\sqrt{\mu^2 - 2g^2 f^2}]
\sim g^2 \frac{f}{\mu} \zeta_{\rm vac} \alpha^{-1} \sim \tilde{c}_b \zeta_{\rm vac} \left(\frac{\omega}{H}\right)^{-1} \sim c_b \left(\frac{\omega}{H}\right)^{-1}.
\ee

\section{Bispectrum properties and saddle point approximation}
\label{sec:saddlepoint}
\begin{figure}
\resizebox{1.0\hsize}{!}{
\includegraphics{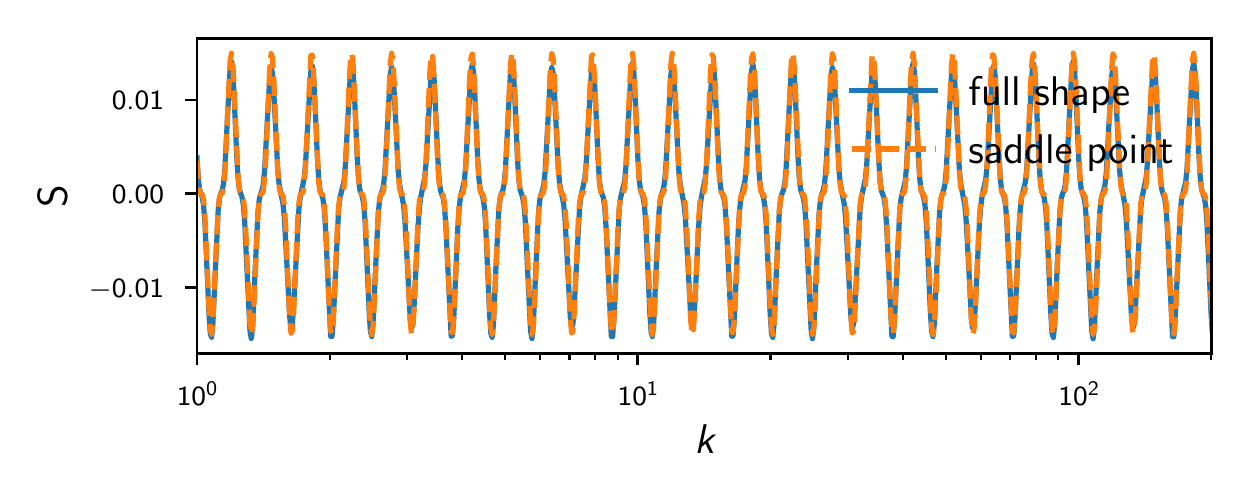}\includegraphics{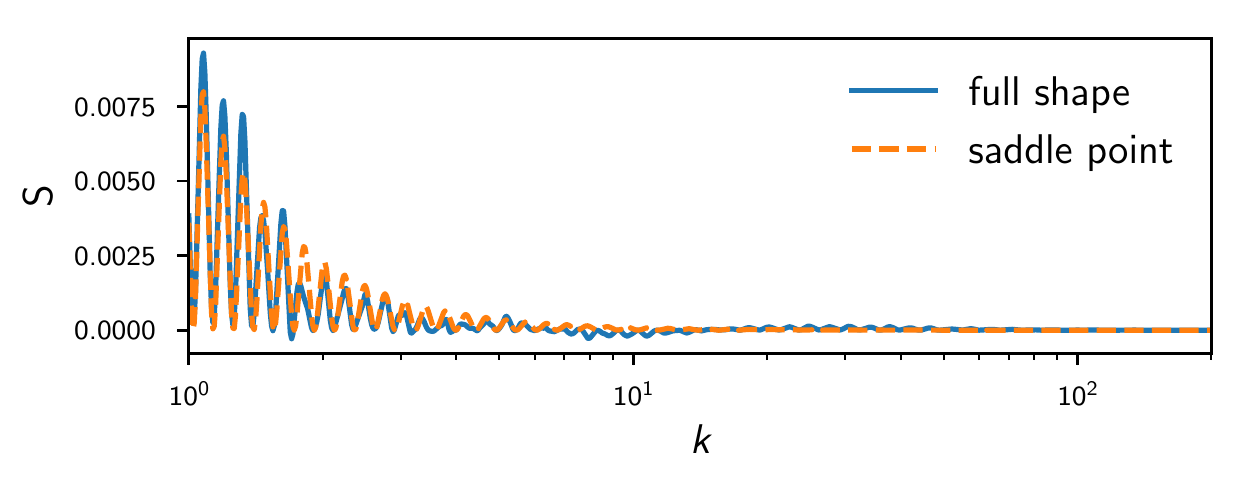}
}
\resizebox{1.0\hsize}{!}{
\includegraphics{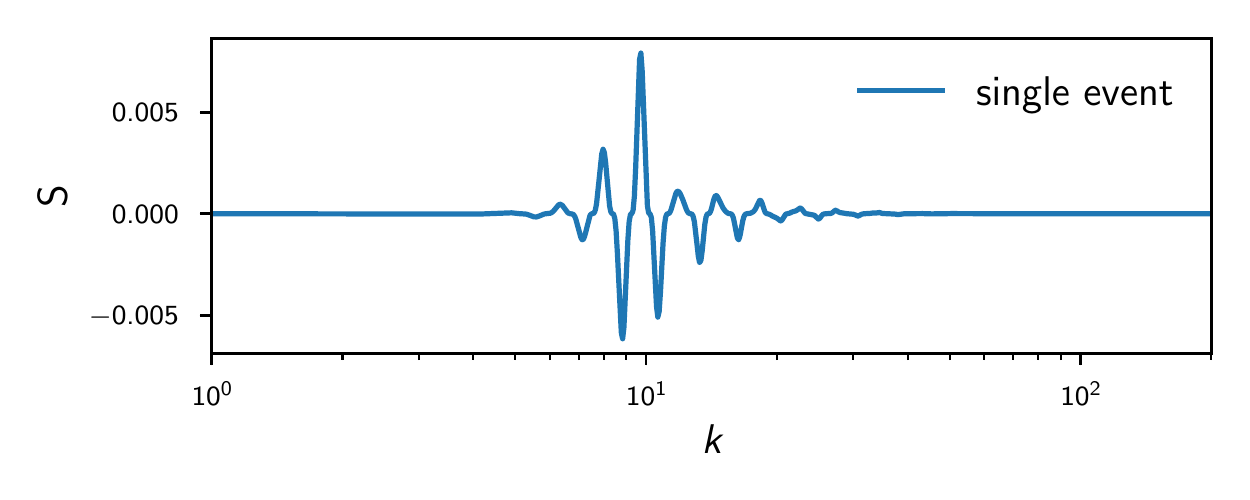}\includegraphics{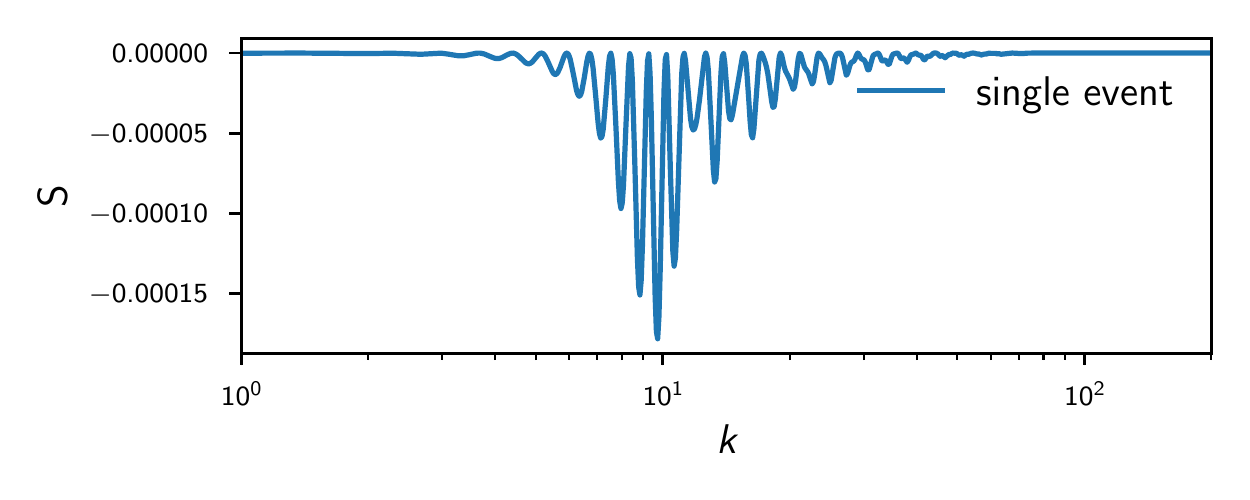}
}
\caption{Top: Comparison of the full shape with the saddle point approximation for $\omega=30$. We plot the shape function of the bispectrum, $S(k_1,k_2,k_3) = (k_1 k_2 k_3)^2 B(k_1,k_2,k_3)$. Left: $k_1=k_2=k_3=k$. Right: $k_1 =1, k_2 = k, k_3 = k$. Bottom: Contribution to the bispectrum from a single production event for the same axes.} 
\label{fig:saddle}
\end{figure}

We illustrate how our shape looks in momentum space at the example of the bispectrum, the most familiar non-Gaussian statistic, in Fig. \ref{fig:saddle} (blue lines), including the sum over production events. The periodic particle creation leads to an oscillating shape, which is therefore orthogonal to the standard bispectrum shapes. The plots also illustrate that the shape is dominated by equilateral contributions. For illustration, we have also plotted the contribution to the bispectrum from a single particle production event. 

The full shape looks very regular and one may wonder whether an approximation can be made to avoid the computationally expensive sum over events. Indeed we have examined if it is possible to remove the sum over events with a saddle point approximation.
We can make a saddle point approximation for $h_n$ (good for large $\omega/H$) giving
\be
\label{shapesaddle8}
h_n(k) \sim \theta \left(-k\eta_n-\frac{\omega}{H} \right)  \, \sin \left(\frac{\omega}{H}\log\frac{k}{H} \right)
\ee
where we dropped some constants and where the $\theta$ function ensures that the saddle is part of the integration range of $\eta$. Plugging this in the shape function we get
\be\label{shapesaddle6}
\langle\zeta_{\k_1}\dots\zeta_{\k_N}\rangle' \sim \sum_n(H\eta_n)^{-3} \theta \left(-k_{min}\eta_n-\frac{\omega}{H} \right)   \prod_{i=1}^N \left( c_b \, \hat{h}(k_i)\right) 
\ee
where we defined the n-independent shape factor
\be
\hat{h}(k) = \frac{1}{k^3} \sin \left( \frac{\omega}{H} \log \frac{k}{H} \right)
\ee
and where $k_{min} = \min(k_1,k_2,...)$. Doing the geometric sum over $\eta_n$ and again dropping constants, we get
\be\label{shapesaddle7}
\langle \zeta_{\k_1}\dots \zeta_{\k_N}\rangle' \sim k_{min}^{3}   \prod_{i=1}^N \left( c_b \, \hat{h}(k_i)\right) 
\ee
This has the correct k scaling for (discretely) scale invariant n-point functions, i.e. $k^{-(3N-3)}$. Unfortunately the minimum condition arising from the $\theta$ functions makes the shape non-factorizable. The bispectrum $B(k_1,k_2,k_3)$ in the saddle point approximation is
\be\label{shapesaddle}
B(k_1,k_2,k_3) = \frac{k_1^{\pm i\omega/H} k_2^{\pm i\omega/H}k_3^{\pm i\omega/H}}{(k_1 k_2 k_3)^3}  k_{min}^3%\left( \max(1/k_1, 1/k_2, 1/k_3) \right)^{-3}
\ee
We show an example of this approximation in Fig.~\ref{fig:saddle} (orange). One might replace the minimum condition with separable forms that select the equilateral range, as this is where most signal lies. However at least for WMAP we have found that we can do the full event sum and thus did not use the saddle point approximation in our analysis.

\end{document}